\documentclass[%
 aip,jcp,superscriptaddress,
 amsmath,amssymb,floatfix,longbibliography,
 reprint,%
]{revtex4-2}
\usepackage{caption}
\usepackage{nicematrix}
\usepackage[utf8]{inputenc}
 \usepackage[T1]{fontenc}
\usepackage{chngcntr}
\usepackage{graphicx}% Include figure files
\usepackage{dcolumn}% Align table columns on decimal point
\usepackage{bm}% bold math
\usepackage{xcolor}
\usepackage[normalem]{ulem}
\usepackage{physics}
\usepackage{chemformula}
\usepackage{soul}
\usepackage{subcaption}
\usepackage{relsize}
\let\ket=\pket %Overwrite physics ket
\usepackage{cancel}
\usepackage{braket}
\usepackage{tikz}
\usetikzlibrary{external}
\usepackage{float}
\usepackage{ragged2e}
\usepackage[colorlinks]{hyperref}
\hypersetup{ % mute the intense default colors for links
    colorlinks=true,
    linkcolor=red!80!black,
    urlcolor=magenta!80!black,
    citecolor=green!70!black,
}
\usepackage{cleveref}
\usepackage{nameref}
\usepackage{ulem} % In your preamble, if not already included

\newcommand{\LiHtwo}{\ensuremath{(\text{LiH})_2}}
\newcommand{\WcmSq}{\text{ W/cm$^2$}}

\DeclareMathOperator*{\argmin}{arg\,min}
\begin{document}
\preprint{}

\title{Time-dependent Gaussian basis sets for many-body systems using Rothe's method: A mean-field study}

\author{Simon Elias Schrader}
\email{s.e.schrader@kjemi.uio.no}
\author{Håkon Emil Kristiansen}%
%\email{haakoek@kjemi.uio.no}

\author{Thomas Bondo Pedersen}%
%\email{t.b.pedersen@kjemi.uio.no}
\author{Simen Kvaal}%
%\email{simen.kvaal@kjemi.uio.no}

\affiliation{Hylleraas Centre for Quantum Molecular Sciences, Department of Chemistry, University of Oslo, P.O. Box 1033 Blindern, N-0315 Oslo, Norway 
}%

\NewDocumentCommand{\removed}{+m}{%
  {\color{red!80!black}#1}%
}
\NewDocumentCommand{\added}{+m}{%
  {\color{blue!80!black}#1}%
}
\date{\today}

\begin{abstract}
A challenge in modeling time-dependent strong-field processes such as high-harmonic generation for many-body systems, is how to effectively represent the electronic continuum. We apply Rothe's method to the time-dependent Hartree-Fock (TDHF) and density functional theory (TDDFT) equations of motion for the orbitals, which reformulates them as an optimization problem. We show that thawed, complex-valued Gaussian basis sets can be propagated efficiently for these orbital-based approaches, removing the need for grids. In particular, we illustrate that qualitatively correct results can often be obtained by using just a few fully flexible Gaussians that describe the unbound dynamics for both TDHF and TDDFT. Grid calculations can be reproduced quantitatively using $30$--$100$ Gaussians for intensities up to $4\times10^{14}\WcmSq$ for the one-dimensional molecular systems considered in this work. 
\end{abstract}

\maketitle

\section{Introduction}\label{sec:Introduction}
Solving the time-dependent Schrödinger equation (TDSE) for molecular systems is necessary to understand, predict and interpret strong-field phenomena such as high-harmonic generation (HHG).\cite{HHG_discovery_1,HHG_discovery_2,Krausz2009,Corkum2007,Haessler2010,Nisoli2017} However, even when the Born-Oppenheimer (BO) approximation is invoked, solving the TDSE generally scales exponentially in the number of particles. A standard approximation to circumvent this exponential scaling is to use orbital-based methods, where the time-dependent wave function is parameterized using a linear combination of antisymmetrized products of one-particle wave functions. Examples of such methods include time-dependent configuration interaction (TDCI),\cite{TDCI} Multiconfigurational time-dependent Hartree-Fock (MCTDHF),~\cite{MCTDHF1,MCTDHF2} which in the limit of just one configuration reduces to time-dependent Hartree-Fock (TDHF), and time-dependent Coupled Cluster (TDCC).\cite{sato2018communication,kvaal2012ab,TDCC_review} In addition to wave-function based methods, time-dependent Kohn-Sham density functional theory (TDDFT) can be used to include correlation effects at a computational cost similar to or even lower than a Hartree-Fock calculation.\cite{RungeGross,TDDFT,Dreuw_HeadGordon,TDDFT_perspective,tancogne2020octopus}

All of these methods have in common that the orbitals need to be able to represent the wave function sufficiently well at all time points, i.e., that a proper basis is chosen. However, isotropic, atom-centered Gaussian functions, developed to represent ground state atoms and molecules,\cite{MEST} are unable to describe unbound, outgoing electrons in the continuum, which occur in strong-field processes such as HHG. While basis sets exist that include continuum contributions,~\cite{Kaufmann,Luppi,Wozniak,Coccia,durden_evaluation_2025} these basis sets often suffer from numerical overcompleteness, they need to be very large, and it is often necessary to make use of heuristic lifetime models to avoid reflections and model ionization.\cite{Klinkusch,Coccia,AbILM} An alternative is to allow the Gaussian basis set to evolve in time, i.e., let the Gaussians have time-dependent (complex) width, position and momentum parameters. In vibrational dynamics, this is done using the variational multiconfigurational Gaussian approach (vMCG).\cite{G-MCTDH,vMCG} However, these approaches struggle with numerical instability due to a non-invertible Gramian matrix appearing in the implicit system of ordinary differential equations, \cite{Sawada,G-MCTDH,vMCG,KAY1989165,Lee2018,HaakonPhD} and different approximations, such as keeping the width parameters time-independent (``frozen''),\cite{Heller_1981,Vanicek2020} are often employed. In addition, the Gramian matrix is usually regularized, which may give rise to large deviations from the real solution,\cite{feischl2024regularized} and it has been demonstrated to be inapplicable for calculations involving atomic systems in strong fields.\cite{HaakonPhD}

{Another approach involving Gaussians are trajectory-guided methods such as the coupled coherent states method (CCS).\cite{Child_HeHe,CCS_H2++,CCS_HHG} In this method, equations of motion are solved for many frozen Gaussians independently. The full quantum dynamics are expressed in a basis consisting of these moving Gaussians. While this method can give excellent results for strong-field dynamics, very large basis sets are required and the correct initial conditions for Gaussians need to be chosen. Furthermore, as those basis sets are not adaptive, basis set incompleteness can become a problem.}

As an alternative to Gaussian-based representations, the wave function or the orbitals can also be represented using a grid or a discrete variable representation (DVR), \cite{Sato_TDCAS_1D,Sato_2016,Sato_TDCC} or using linear basis functions with compact support, such as B-splines.~\cite{Bspline_1,Bspline_2,Bspline_3} However, these methods scale extremely steeply in the number of particles when correlation is taken into account, and require very large bases or a large number of grid points. Gaussian basis sets, on the other hand, give a very compact representation of the wave function, and it is desirable to represent the wave function using such a basis. Recently, Rothe's method has been used to overcome the numerical problems that arise in the vMCG method by reformulating time evolution as an optimization problem.\cite{kvaal2023need,schrader2024time,schrader2025HeHe,wozniak2025rothetimepropagationcoupled} At each time point, a new set of Gaussians is found that optimally represents the wave function at that time point, bypassing the need to solve very stiff ordinary differential equations and also allowing for a flexible representation of the wave function by having the ability to add or remove basis functions (which, however, is also possible with recent advances in the vMCG method, see Ref. \citenum{Martinazzo_Burghardt}).
It has been demonstrated that Rothe's method can be used to propagate the three-dimensional (3D) electronic wave function for the hydrogen atom in a strong electric field, and that it can be well represented using only a few dozen Gaussian wave packets at all time points. Explicitly correlated Gaussians, which give a very compact representation of the time-dependent wave function, can also be propagated effectively. However, previous formulations of Rothe's method for the Schrödinger equation have required that the full Schrödinger equation be solved approximately, not just the orbital equations that arise in many methods.

In this work, we present a novel application of Rothe's method to orbital-based time-dependent methods, specifically TDHF and TDDFT. We exemplify this by studying the dynamics of one-dimensional (1D) molecular model systems in strong fields. While 1D systems are unable to capture all aspects of realistic 3D molecular systems, they serve as valuable test systems for evaluating numerical methods. Due to the numerical tractability of 1D systems, it is possible to perform numerically exact reference calculations even for complex many-body systems, enabling rigorous benchmarking. Furthermore, many of the numerical challenges present in 3D calculations, such as basis set convergence and propagation stability, can be similar in 1D, making insights gained from these models transferable to realistic systems. Hence, 1D systems have been used to study different numerical schemes for propagating 1D molecules in strong fields.\cite{Sato_TDCAS_1D} Similarly, exchange-correlation functionals have been developed for and applied to 1D systems.\cite{Helbig_1DLDA,1DDFT_application_Wagner,1DDFT_construction,1DDFT_exponentials,1DDFT_exact} Our results demonstrate that the application of Rothe's method to orbital-based approaches provides a compact and numerically stable representation of electronic dynamics in strong fields. In Sec.~\ref{sec:Methods}, we describe the theory behind TDHF and TDDFT, how Rothe's method can be applied to orbital equations, and how a good basis representation for the initial state can be obtained. This is followed by a description of the systems and functionals under consideration in Sec.~\ref{sec:Systems}. In Sec.~\ref{sec:Numerics}, we describe the numerical details of the propagation, in addition to a discussion of the reference calculations. We present and discuss our results in Secs. \ref{sec:Results} and \ref{sec:Discussion}, respectively. We conclude with a summary of our findings and future perspectives in Sec.~\ref{sec:Conclusion}. Unless stated otherwise, atomic units (a.u.) are used in this work. 

\section{Theory}\label{sec:Methods}
\subsection{TDHF and TDDFT}
In Hartree-Fock theory (HF)\cite{Szabo&Ostlund} and Kohn-Sham Density Functional Theory (KS-DFT),\cite{KohnShamDFT} an $N$-electron wave function $\Psi(\boldsymbol{x}_1,\dots,\boldsymbol{x}_N)$ is represented as a single Slater determinant using a set of $N$ orthonormal spin orbitals $\varphi_j(\boldsymbol{x})$:
\begin{equation}\label{eq:SlaterDeterminant}
    \Psi(\boldsymbol{x}_1,\dots,\boldsymbol{x}_N)=\frac{1}{\sqrt{N!}}\hat{A} \left(\varphi_1(\boldsymbol x_1)\dots\varphi_N(\boldsymbol x_N)\right),
\end{equation}
with the associated density
\begin{equation}\label{eq:Density}
    \rho(\boldsymbol r) = N \int |\Psi(\boldsymbol{r}, s_1, \boldsymbol x_2, \ldots, \boldsymbol x_N)|^2 \, \mathrm{d}s_1 \mathrm{d}\boldsymbol x_2 \cdots \mathrm{d}\boldsymbol x_N,
\end{equation}
where $\hat{A}$ is the antisymmetrization operator
and $\boldsymbol{x}_i={\{\boldsymbol{r}_i,s_i\}}$ is a collective variable that contains spin $s_i$ and position $\boldsymbol{r}_i$ of each electron. In TDHF theory\cite{Dirac1930,Dreuw_HeadGordon} and TDDFT,\cite{RungeGross,Dreuw_HeadGordon} the wave function and the density are still represented according to eqs. \eqref{eq:SlaterDeterminant} and \eqref{eq:Density} but the spin orbitals are now time dependent. Omitting the dependency on $\boldsymbol{x}$ for the spin orbitals for notational convenience, the equations that govern the time evolution of each spin orbital $\varphi_j(t)$ read
\begin{equation}\label{eq:TDHFDFT}
    \frac{\partial}{\partial t} \varphi_j=-i\hat{h}(\Phi(t),t) \varphi_j,
\end{equation}
where $\Phi(t)=\{\varphi_i(t)\}_{i=1}^N$ is a set of time-dependent spin orbitals that make up the wave function and density.

In the case of TDHF,
\begin{equation}\label{eq:FockOperator}
\hat{h}(\Phi(t),t)=\hat{F}(\Phi(t),t) 
\end{equation}
where $\hat{F}(\Phi(t),t)$
is the Fock operator,
\begin{equation}
    \hat{F}(\Phi(t),t)=-\frac{1}{2}\nabla^2+v_{\rm {ext}}(\boldsymbol{r},t)+J(\Phi(t))-K(\Phi(t)),
\end{equation}
where $v_{\rm {ext}}(\boldsymbol{r},t)$ is a potentially time-dependent potential describing the interaction between the electron and the nuclei as well as time-dependent field terms, while $\hat{J}(\Phi(t))$ and $\hat{K}(\Phi(t))$ stand for the direct and the exchange operator, respectively:
\begin{equation}
\hat{J} \varphi_j(\mathbf{x}) = \sum_{i=1}^N\left[ \int {|\varphi_i(\mathbf{x}')|^2}w(\boldsymbol{x},\boldsymbol{x}') \, \text{d}\mathbf{x}' \right] \varphi_j(\mathbf{x})\end{equation}
\begin{equation}
  \hat{K} \varphi_j(\mathbf{x}) = \sum_{i=1}^N\left(\left[ \int {\varphi_i^*(\mathbf{x}') \varphi_j(\mathbf{x}')}w(\boldsymbol{x},\boldsymbol{x}') \, \text{d}\mathbf{x}'\right]  \varphi_i(\mathbf{x}) \right).
\end{equation}
For notational convenience, we here omitted the dependence of $\hat{J}$ and $\hat{K}$ on the spin orbitals. The symbol $w(\boldsymbol{x},\boldsymbol{x}')$ denotes the Coulombic electron-electron repulsion.

In the case of TDDFT, the time-dependent wavefunction is not the physical wavefunction, but rather the uncorrelated wavefunction of a fictitious non-interacting system with the same density as the physical system. Using the adiabatic approximation, the spin-orbitals of the fictitious system evolves according to\cite{Gross_Kohn}
\begin{equation}\label{eq:TDDFTOperator}
\hat{h}(\Phi(t),t)=-\frac{1}{2}\nabla^2+v_s(\Phi(t),t),
\end{equation}
where ${ v_{s}(\Phi(t),t)}$ is the Kohn-Sham potential,
\begin{equation}
    { v_{s}(\Phi(t),t)=v_{\rm {ext}}(\boldsymbol{r},t)+v_{J}(\Phi(t))+v_{\rm {XC}}(\Phi(t))}.
\end{equation}
Here, $v_J(\Phi(t))=\hat J(\Phi(t))$ is the Hartree (direct) potential and $v_{\rm {XC}}(\Phi(t))$ the exchange-correlation (XC) potential. In KS-DFT, it is obtained as the functional derivative of the XC functional $E_{\rm XC}[\rho]$ with respect to the density,
\begin{equation}
    v_{\rm {XC}}(\rho)=\frac{\delta E_{\rm XC}[\rho]}{\delta \rho(\boldsymbol{r})}.
\end{equation}
It should be noted that $v_J$ and $v_{\rm {XC}}$ are formally functions of the time-dependent density $\rho(\boldsymbol{r},t)$ only, not the spin orbitals. However, we choose to write them as functions of the time-dependent spin orbitals, as depending on the choice of the XC potential, spin orbital information such as kinetic energy density may enter, and also to keep the similarity to Hartree-Fock theory apparent. This notation is consistent, as the density is ultimately calculated from the spin orbitals. The choice of the XC functional in this work is discussed in Sec.~\ref{sec:1DXC}. While the equations presented here are for a general framework, we used a spin-restricted implementation of HF and KS-DFT in this work, i.e., each spatial orbital is doubly occupied by one spin-up and one spin-down electron, and all molecular systems considered are closed-shell systems. As this removes spin-dependency from the equations, we will hereafter only refer to orbitals.
{
\subsection{Variable Projection (VarPro)}\label{sec:VarPro}
We consider here a fitting problem of the type 
\begin{equation}\label{eq:lsq}
\varepsilon(\boldsymbol{c},\boldsymbol{\alpha})=\norm{\sum_{m=1}^M c_m g_m(\boldsymbol{\alpha})-f}^2,
\end{equation}

\begin{equation}\label{eq:Fitting}
\boldsymbol{c}_{\mathrm {opt.}},\boldsymbol{\alpha}_{\mathrm {opt.}}=\argmin_{\boldsymbol{c},\boldsymbol{\alpha}}\varepsilon(\boldsymbol{c},\boldsymbol{\alpha}),
\end{equation}

where $\{g_m(\boldsymbol{\alpha})\}_{m=1}^M$ is a set of functions that depend nonlinearly on parameters $\boldsymbol{\alpha}$, $\boldsymbol{c}$ is a set of linear coefficients, and $f$ is a target function. The goal is hence to find the coefficients $\boldsymbol{c}_{\mathrm {opt.}},\boldsymbol{\alpha}_{\mathrm {opt.}}$ that minimize the fitting error $\varepsilon$ in an $L^2$ sense.
The \emph{Variable Projection} (VarPro) method simplifies this to an optimization over the nonlinear coefficients $\boldsymbol{\alpha}$ only, while the linear coefficients are obtained analytically for a given $\boldsymbol{\alpha}$. \cite{Golub_Pereyra,OLeary2013}

We consider here two cases, one where the problem is solved in Hilbert space directly, and one where it is solved on a grid.  In Hilbert space, we define the matrix $\boldsymbol{S}(\boldsymbol{\alpha})$ with elements
\begin{equation}
    \boldsymbol{S}_{m,n}(\boldsymbol{\alpha})=\langle g_m(\boldsymbol{\alpha})|g_n(\boldsymbol{\alpha})\rangle,
\end{equation}
and the vector $\boldsymbol{\rho}$ with elements
\begin{equation}
    \boldsymbol{\rho}_m(\boldsymbol{\alpha})=\langle g_m(\boldsymbol{\alpha})|f\rangle.
\end{equation}
Then, for given nonlinear coefficients $\boldsymbol{\alpha}$, the linear coefficients $\boldsymbol{c}$ that minimize Eq.~\ref{eq:lsq} are obtained by
\begin{equation}
    \boldsymbol{c}(\boldsymbol{\alpha})=\boldsymbol{S}(\boldsymbol{\alpha})^{-1}\boldsymbol{\rho}(\boldsymbol{\alpha}),
\end{equation}
and the $L^2$ difference $\varepsilon$ to be optimized can be considered a function of only the nonlinear coefficients,
\begin{equation}
    \varepsilon(\boldsymbol{\alpha})\equiv\varepsilon(\boldsymbol{c}(\boldsymbol{\alpha}),\boldsymbol{\alpha}).
\end{equation}
If $\boldsymbol{S}(\boldsymbol{\alpha})$ and $\boldsymbol{\rho}(\boldsymbol{\alpha})$ are differentiable with respect to $\boldsymbol{\alpha}$, analytical gradients of $\varepsilon(\boldsymbol{\alpha})$ exist. In the case of (numerical) linear dependency, Tikhonov regularization can be used, i.e. one replaces $\boldsymbol{S}(\boldsymbol{\alpha})^{-1}$ with $\left(\boldsymbol{S}(\boldsymbol{\alpha})+\lambda \boldsymbol{I}\right)^{-1}$ for a small value $\lambda>0$, where $\boldsymbol{I}$ is the $M\times M$ identity matrix.

We now consider the grid case. Let $\{x_k\}_{k=1}^K$ be a set of (possibly multi-dimensional) quadrature points with corresponding quadrature weights $w_k$. We define the $K\times M$ matrix $\Phi(\boldsymbol{\alpha})$ with elements 
\begin{equation}
\Phi_{k,m}(\boldsymbol{\alpha})=g_m(\boldsymbol{\alpha})(x_k),
\end{equation}
and the vector $\boldsymbol{f}$ with elements
\begin{equation}
    \boldsymbol{f}_k=f(x_k).
\end{equation}
The fitting error then reads
\begin{equation}\label{eq:Quadrature}
    \varepsilon(\boldsymbol{c},\boldsymbol{\alpha})=\norm{\boldsymbol{W}\left(\boldsymbol{f}-\Phi \boldsymbol{c}\right)}^2,
\end{equation}
where $\boldsymbol{W}$ is a diagonal matrix with elements $\boldsymbol{W}_{k,k}=\sqrt{w_k}$. Equation \eqref{eq:Quadrature} is the quadrature approximation of the Hilbert space inner product. The analytical solution for $\boldsymbol{c}(\boldsymbol{\alpha})$ is now given by
\begin{equation}\label{eq:GridVarPro}
    \boldsymbol{c}(\boldsymbol{\alpha)}=(\boldsymbol{W}\Phi)^{+}\boldsymbol{W}\boldsymbol{f},
\end{equation}
where $(\boldsymbol{W}\Phi)^{+}$ is the Moore-Penrose pseudoinverse of $\boldsymbol{W}\Phi$. Observe that in general $(\boldsymbol{W}\Phi)^{+}\neq\Phi^+\boldsymbol W^{+}$, hence $\boldsymbol{W}$ does not cancel out. Numerically, Eq.~\eqref{eq:GridVarPro} can be implemented in a stable way using the singular value decomposition (SVD), where a threshold can be used on the smallest singular values in order to avoid numerical instabilities due to linear dependency. Analytical derivatives for $\varepsilon(\boldsymbol{\alpha})$ exist when $\{g_m(\boldsymbol{\alpha})\}_{m=1}^M$ is differentiable with respect to $\boldsymbol{\alpha}$, see Ref.~\citenum{OLeary2013} for details regarding the concrete implementation of how to analytically differentiate through the SVD.
}

\subsection{Rothe's method for orbital equations}\label{sec:RothesMethod}
We introduce here the working equations for the application of Rothe's method to the TDHF and TDDFT equations. We refer to Refs. \citenum{kvaal2023need,schrader2024time} for a derivation of Rothe's method for the time-dependent Schrödinger equation with explicitly correlated Gaussian functions and to Ref. \citenum{schrader2025HeHe} for such an application.

Starting from Eq.~\eqref{eq:TDHFDFT}, we let the operator $\hat{h}(\Phi,t)$ stand for either the Fock operator (Eq.~\eqref{eq:FockOperator}) or the DFT propagation operator (Eq.~\eqref{eq:TDDFTOperator}). We introduce a semi-discretization of Eq.~\eqref{eq:TDHFDFT} by first assuming that over a time interval $t \in [t_i, t_i+\Delta t]$, we have $\hat{h}(\Phi(t),t)\approx \hat{h}(\Phi(t_i),t)$, thereby assuming that the orbital-dependent terms
of $\hat{h}$ remain approximately constant, and next by using the Crank--Nicolson propagator over the time interval. This scheme is a constant-mean field (CMF) integrator, related to but distinct from the second-order CMF integration scheme common in the MCTDHF community.~\cite{MCTDHbook} Our CMF scheme has global error $O(\Delta t)$, but we observe excellent stability and qualitative behavior despite its low order. We define the operator
\begin{equation}
    \tilde A_{i}(\Phi(t_i),h)=\hat I+\text{i}\frac{\Delta t}{2}\hat h\left(\Phi(t_i),t_i+\frac{\Delta t}{2}\right),
\end{equation}
and obtain the $j$'th orbital at the next time point $t_{i+1}=t_i+\Delta t$ as
\begin{equation}
    \varphi_j(t_{i+1})=\tilde A_{i}^{-1} \tilde A_{i}^\dagger\varphi_j\left(t_i\right),
\end{equation}
where we omitted the dependence of $\tilde{A}$ on $t_i$, $\Phi(t_i)$ and $\Delta t$ for brevity. Thus, we have for each orbital $\varphi_j$ that
\begin{equation}
    \norm{\tilde A_{i}\varphi_j(t_{i+1})- \tilde A_{i}^\dagger\varphi_j\left(t_i\right)}^2=0.
\end{equation}
As this holds true for all orbitals, we also have that  
\begin{equation}
    \sum_{j=1}^N\norm{\tilde A_{i}\varphi_j(t_{i+1})- \tilde A_{i}^\dagger\varphi_j\left(t_i\right)}^2=0
\end{equation}
or equally
\begin{equation}
    \Phi(t_{i+1})=\argmin_{\Omega}\sum_{j=1}^N\norm{\tilde A_{i}\omega_j- \tilde A_{i}^\dagger\varphi_j\left(t_i\right)}^2
\end{equation}
where $\Omega=\{\omega_i\}_{i=1}^N$, and the notation $\argmin_{\Omega}$ means that optimization is carried out for all orbitals $\omega_j$, $j=1,\dots,N$. 

Each $\omega_j$ is now expanded using $M$ basis functions $\left\{g_m(\boldsymbol{\alpha})\right\}_{m=1}^M$ that depend in a nonlinear fashion on a parameter vector $\boldsymbol{\alpha}$. Hence, we write
\begin{equation}
  \omega_j=\sum_m^M c_{j,m}g_m(\boldsymbol{\alpha}).  
\end{equation}
Introducing the \textit{orbital Rothe error} $r_{i+1}^j$ as the error in orbital $j$ going from time point $i$ to $i+1$, 
\begin{equation}
r_{i+1}^j(\boldsymbol{c},\boldsymbol{\alpha})=\norm{\sum_{m=1}^{M} c_{j,m}\tilde A_{i}g_m(\boldsymbol{\alpha})- \tilde A_{i}^\dagger\varphi_j\left(t_i\right)}^2    
\end{equation}
and the all-orbital Rothe error (or simply \textit{Rothe error})
\begin{equation}\label{eq:RotheError}
    r_{i+1}(\boldsymbol{c},\boldsymbol{\alpha})=\sqrt{\sum_{j=1}^N r_{i+1}^j(\boldsymbol{c},\boldsymbol{\alpha})},
\end{equation}
the optimal parameters at time $t_{i+1}$ are given by 
\begin{equation}\label{eq:Rothe_opt}
    \boldsymbol{\alpha}(t_{i+1}),\boldsymbol{c}(t_{i+1})=\argmin_{\boldsymbol{\alpha},\boldsymbol{c}}r_{i+1}(\boldsymbol{c},\boldsymbol{\alpha}).
\end{equation}
This gives the orbitals at the next time point
\begin{equation}
    \varphi_j(t_{i+1})=\sum_m^M c_{j,m}(t_{i+1})g_m(\boldsymbol{\alpha}(t_{i+1})).
\end{equation}
The square root of the orbital Rothe error $r^j_{i+1}(\boldsymbol{c}(t_{i+1}),\boldsymbol{\alpha}(t_{i+1}))$ is an upper bound for the time evolution error in orbital $\varphi_j$ going from time $t_i$ to $t_{i+1}$ (relative to the chosen integrator, i.e., relative to the Crank-Nicolson propagation with the constant mean field approximation).
Unless each $r^j_{i+1}(\boldsymbol{c}(t_{i+1}),\boldsymbol{\alpha}(t_{i+1}))$ is zero for each $j=1,\dots,N$, which requires the basis representation of the orbitals to be exact, Crank-Nicolson propagation will only be approximated. We also define the cumulative Rothe error at the final time of a calculation as the sum over all Rothe errors going from time $t_0$ to the final time $t_f$ using $N_T=(t_f-t_0)/h$ time points
\begin{equation}\label{r_tot}
 r_{\text{tot.}}=\sum_{i=1}^{N_T}r_{i}(\boldsymbol{\alpha}(t_i),\boldsymbol{c}(t_i)).
\end{equation}
One can consider $r_{\text{tot.}}$ to be an overall measure of the quality of the Rothe propagation. It should however be noted that it is not an explicit upper bound for the time evolution error, as it does not take into account the coupling between the orbitals. 
While the time propagation equations \eqref{eq:TDHFDFT} conserve orthonormality (and energy for time-independent Hamiltonians), this is not guaranteed with Rothe's method, as Crank-Nicolson propagation is only approximated.
A scheme that enforces orthonormality is discussed in Sec.~\ref{sec:conservationrules}.

From a numerical point of view, no iterative optimization needs to be carried out for the linear coefficients $\boldsymbol{c}(t_{i+1})$. As orthonormalization is not explicitly enforced, the optimization of the linear coefficients in Eq.~\eqref{eq:Rothe_opt} for each orbital $j$ does not depend on the other orbitals. Thus, for each orbital $j$, the optimization of the linear coefficients $c_{*,m}$ for a given set of nonlinear coefficients $\boldsymbol{\alpha}$ can be carried out using the {VarPro} method, as was done in previous applications of Rothe's method for the Schrödinger equation. 
\cite{kvaal2023need,schrader2024time,schrader2025HeHe} Hence, explicit optimization is only carried out with respect to the nonlinear coefficients $\boldsymbol{\alpha}$ and, therefore, we write the Rothe error as a function of $\boldsymbol{\alpha}$ only. 

In order to ensure that the correct mean-field dynamics are properly reproduced, we introduce a parameter $\varepsilon_{\Delta t}/N_T$ that controls the deviation from the exact dynamics: If the Rothe error (after optimization) is greater than $\varepsilon_{\Delta t}/N_T$, it indicates that the underlying set of basis functions is no longer able to describe the dynamics properly. In that case, we enlarge the basis by adding one more basis function, a procedure described in Sec.~\ref{sec:AddGaussians}. This guarantees that the cumulative Rothe error is at most $\varepsilon_{\Delta t}$. {Thus, by choosing a sufficiently small $\varepsilon_{\Delta t}$, any process can be in principle be described with very high precision whenever a basis set is used that allows for completeness.}

\subsection{Gaussian basis functions and ground state orbitals}\label{sec:InitOrbs}

As in our previous applications of Rothe's method, we use Gaussian functions as basis functions. Each orbital $\varphi_j$, $j=1,\dots,M$, is represented as a linear combination of $M$ Gaussians of the form 
\begin{align}
    &g_m(\boldsymbol{\alpha}) = g(\boldsymbol{\alpha}_m) \nonumber \\
    \label{eq:Gaussian}
    &= d_m\exp\left(-(a_m^2+\text{i}b_m)\left({x-{\mu}_m}\right)^2+\text{i}{p}_m(x-{\mu}_m)\right),
\end{align}
where $\boldsymbol{\alpha}_m=\{a_m,b_m,\mu_m,p_m\}$ is a set of real numbers, and $d_m$ is the real, nonnegative number that ensures normalization of the Gaussian, i.e.,
\begin{equation}
    d_m=\left({\frac{2a_m^2}{\pi}}\right)^{1/4}.
\end{equation}

The set of parameters $\boldsymbol{\alpha}=\{\boldsymbol{\alpha}_m\}_{m=1}^M$ represents the nonlinear coefficients of all $M$ Gaussians.  
The $j$th orbital then reads
\begin{equation}
    \varphi_j=\sum_{m=1}^M c_{j,m}g_m(\boldsymbol{\alpha}).
\end{equation}

Unlike 3D molecular and atomic systems, no optimized Gaussian basis sets are available for 1D systems, and special care must be taken to find a set of Gaussian functions that is well suited to represent the ground state.

In order to obtain a starting guess for the orbitals that define the ground state, we implemented a two-step procedure. First, we perform a grid calculation as described in Sec.~\ref{sec:Grid}, giving us a set of $N$ occupied orbitals $\{\varphi^G_j\}_{j=1}^N$, where the superscript 'G' stands for grid. We then fit a given number of Gaussians $M$ to the $N$ orbitals, i.e., for a given number of Gaussians $M$, we obtain a set of linear coefficients $c_{j,m}$ and nonlinear coefficients $\boldsymbol{\alpha}$ that minimize the residual
\begin{equation}
\varepsilon({\boldsymbol{c},}\boldsymbol{\alpha})=\sum_{j=1}^{N} \varepsilon^j({\boldsymbol{c}_{j},}\boldsymbol{\alpha})
\end{equation}
where 
\begin{equation}\label{eq:Eq26}
    \varepsilon^j({\boldsymbol{c}_{j},}\boldsymbol{\alpha})=\norm{\sum_{m=1}^M c_{j,m}g_m(\boldsymbol{\alpha})-\varphi^G_j}^2.
\end{equation}
As in Rothe's method, the optimal linear coefficients can be obtained analytically for a given set of nonlinear coefficients by means of least squares using the VarPro algorithm, which is applicable to each orbital independently. Hence, the total residual $\varepsilon$ is a function of the nonlinear coefficients $\boldsymbol{\alpha}$ only, since the optimal linear coefficients ${\boldsymbol{c}}$ can be parametrized as functions of $\boldsymbol{\alpha}$. Optimization of the residual $\varepsilon$ is then carried out using the Broyden-Fletcher-Goldfarb-Shanno (BFGS) algorithm\cite{nocedal_numerical} with analytical gradients. The number of Gaussians used is system dependent, see Sec. \ref{sec:InitOrbitals}.
This gives us a set of orbitals $\{\varphi_j\}_{j=1}^N$ written as a linear combination of Gaussians. Although slightly more involved, this approach is more efficient than fitting Gaussians to each orbital individually, as it is invariant under unitary transformations of the orbitals {$\{\varphi_j\}_{j=1}^N$} and leads to a minimal number of Gaussians. This approach also reduces the likelihood that Gaussians have a large overlap, as accidental overlaps between Gaussians are avoided. Although real Gaussians suffice for the ground-state wavefunction, we did not restrict the Gaussians to be real-valued, as complex-valued Gaussians are anyway used for the subsequent time evolution.

However, this approach does not guarantee orthonormality between the orbitals $\{\varphi_j\}_{j=1}^N$, and it does not guarantee an optimally small initial Rothe error for a given amount of Gaussians either. Hence, as the second step, we proceed to use imaginary time propagation using Rothe's method to further optimize the linear and nonlinear coefficients. That is, we use Crank-Nicolson propagation as described in Sec. \ref{sec:RothesMethod} using the time step $-i\Delta t$ where $\Delta t= 0.05$. Thereby, we further optimize the linear and nonlinear coefficients that represent the orbitals. This procedure leads to a further reduction of the energy, guarantees orthonormal orbitals, and gives rise to an optionally small initial Rothe error. The initial Rothe error $r_1$, i.e. the Rothe error going from time point $t_0=0$ to time point $t_1=\Delta t$ when no external field is active, is the error that arises because the ground state isn't represented exactly. Getting this error as low as possible within a given basis avoids spurious oscillations in the time propagation, and makes sure that the Rothe error represents a change in the wave function due to an external field inducing non-stationarity, not due to an insufficient ground state.

\subsection{Alternative Rothe's method for TDHF}

An alternative approach is to propagate a single Slater determinant according to the TDSE with the constraint that the wave function remains a Slater determinant, i.e.,
\begin{align}
    \Psi(t_{i+1})=&\argmin_\Omega \norm{\hat{A}_i\Omega-\hat{A}_i^\dagger\Psi(t_i)}^2 \\
    &\text{s.t. $\Omega$ is a normalized Slater determinant}, \nonumber
\end{align}
where 
\begin{equation}
    \hat A_{i}=\hat I+\text{i}\frac{\Delta t}{2}\hat H\left(t_i+\frac{\Delta t}{2}\right).
\end{equation}
In this approach, the Fock operator needs not be constructed.
The optimization is again carried out with respect to the nonlinear coefficients of each basis function $\boldsymbol{\alpha}$ and the linear coefficients of each basis function $c_{j,m}$, i.e., we write 
\begin{equation}\label{eq:Rothe_alt}
\Psi(t_{i+1})=\argmin_\Omega \norm{\hat{A}_i\ket{\omega_1,\dots,\omega_N}-\hat{A}_i^\dagger\Psi(t_i)}^2,
 \end{equation}
where each $\omega_j$ is again expanded as
$\omega_j=\sum_m^M c_{j,m}g_m(\boldsymbol{\alpha})$, and $\ket{\omega_1,\dots,\omega_N}$ is a Slater determinant composed of the orbitals $\{\omega_j\}_{j=1}^N$. The difference between this approach (approach 2) and the approach described in the previous section (approach 1), is that approach 1 can give a Rothe error that is arbitrarily close to $0$, as we solve evolution equations for the orbitals which are derived using the time-dependent variational principle,\cite{broeckhove_equivalence_1988,Lasser_TDVP} which can in principle be approximated to arbitrary accuracy using a complete basis set such as Gaussians.

However, approach 2 might not give small Rothe errors, as there will be an irreducible error: the propagation of a Slater determinant under the exact Hamiltonian will generally not remain a Slater determinant. In this work, we have not pursued approach 2 further, as we implemented the evaluation of the Rothe error on a grid {(see Sec.~\ref{sec:Evaluation} for details)}. This simplifies the calculation for approach 1, as one circumvents the implementation of matrix elements of the squared Fock operator $\hat{F}^2$. However, for an $N$-orbital Slater determinant, the calculation of \cref{eq:Rothe_alt} on a grid requires to store an $N$-dimensional function in memory and is hence not feasible even for very small systems. We would, however, like to stress that this approach might easily be tested if matrix elements of $\hat{H}^2$ are implemented: The calculation of \cref{eq:Rothe_alt} requires the use of generalized Slater-Condon rules \cite{mayer_simple_2003,GSC} for three-body interactions in addition to matrix elements between up to three-particle functions: Namely, expectation values of the terms $V(i,j)V(i,k)$ are required. Most of those are also required for $\hat{F}^2$.

\section{Model systems}\label{sec:Systems}

\subsection{One-dimensional test systems}\label{sec:Molecules}

Following \citeauthor{Sato_TDCAS_1D},\cite{Sato_TDCAS_1D} we here study lithium hydride (LiH) and the lithium hydride dimer \LiHtwo\ in a strong laser field. The 1D Hamiltonian in the clamped-nuclei Born-Oppenheimer approximation with a softened Coulomb potential, using the electric dipole approximation, reads
\begin{equation}\label{eq:Hamiltonian}
\begin{aligned}
\hat H= & \mathlarger{\sum}_i^{N}\left(-\frac{1}{2} \frac{\partial^2}{\partial x_i^2}-\mathlarger{\sum}_a^{N_a} W(x_i, X_a; Z_a, c) +E(t) x_i\right) \\
& +\mathlarger{\sum}_{i>j}^{N} W(x_i, x_j; 1, d)+\mathlarger{\sum}_{a>b}^{N_a} W(X_a, X_b; 1, 0),
\end{aligned}    
\end{equation}
where $N$ stands for the number of electrons, $x_i$ for the position of electron $i$, and $N_a$ stands for the number of nuclei, with $X_a$ and $Z_a$ representing the charge and position of nucleus $a$, respectively, and $W(x_1, x_2; Z, c)$ is the softened Coulomb potential 
\begin{equation}
    W(x_1, x_2; Z, c) \equiv \frac{Z}{\sqrt{(x_1-x_2)^2+c}} \label{eq:soft_coulomb_def},
\end{equation}
where $c$ is referred to as the damping parameter.

For the damping parameters $c$ and $d$, we use $c=0.5$ and $d=1$. For LiH, the charges read $\boldsymbol{Z}=\{3,1\}$, and the positions are $\boldsymbol{X}=\{-1.15,1.15\}$. For \LiHtwo, we set $\boldsymbol{Z}=\{3,1,3,1\}$ and $\boldsymbol{X}=\{-4.05,-1.75,+1.75,+4.05\}$. For the external time-dependent electric field, we set 
\begin{equation}
    E(t)=-E_0f(t)\sin(\omega t),
\end{equation}
where $f(t)$ is the trigonometric envelope,\cite{Barth2009}
\begin{equation}
  f(t) = 
  \begin{cases} 
    \sin^2\!\left(\frac{\pi t}{t_f}\right), & \text{if } 0 \leq t \leq t_f, \\ 
    0, & \text{otherwise},
  \end{cases}
\end{equation}
where $t_f = 2\pi N_c / \omega$ with $N_c$ the number of optical cycles. For all simulations, we set $N_c=3$ and $\omega=0.06075$ a.u.~corresponding to a wavelength of $\lambda=750$ nm. We consider strong fields with field strength $E_0=0.0534$ a.u.~(corresponding to the peak intensity $I_0=10^{14} \WcmSq$) and $E_0=0.1068$ a.u.~($I_0=4\times10^{14} \WcmSq$). All these parameters are consistent with those used in Ref. \citenum{Sato_TDCAS_1D}.

\subsection{One-dimensional XC functionals}\label{sec:1DXC}

Local density approximation (LDA) functionals have been designed for 1D electronic systems interacting through the soft Coulomb potential with unit dampening
parameter (i.e., with $d=1$ as above).\cite{Helbig_1DLDA,1DDFT_application_Wagner}
Using the 1D LDA, the XC potential $E_{\rm XC}(\rho)$ is a functional of the density $\rho(x)$ 
\begin{equation}
    E_{\rm XC}^{\rm LDA}[\rho] = \int \rho(x) \varepsilon_{\rm XC}(\rho(x)){\rm d}x,
\end{equation}
where $\varepsilon_{\rm XC}$ is the XC energy density per particle. We separate $\varepsilon_{\rm XC}$ into exchange and a correlation contributions
\begin{equation}
    \varepsilon_{\rm XC}(\rho(x))=\varepsilon_{\rm X}^{\rm unif.}(\rho(x))+\varepsilon_{\rm C}(\rho(x)),
\end{equation}
where the exchange contribution $\varepsilon_{\rm X}^{\rm unif.}$ is derived from a 1D uniform electron gas,\cite{1DDFT_application_Wagner}
\begin{equation}
\varepsilon_{\mathrm{X}}^{\mathrm{unif.}}(\rho) = -\frac{\rho}{2}\int_0^\infty \, \frac{\sin^2 y}{y^2\sqrt{\left(\pi \rho / 2\right)^2 + y^2}} \mathrm{d}y.
\end{equation}
To avoid solving this integral repeatedly for all densities $\rho(x)$, we use a cubic spline fit to this expression, making sure that the absolute error is below $10^{-6}$ in the density range $\rho\in[10^{-12},10^4]$, while using asymptotic expressions outside of that region. The correlation contribution $\varepsilon_{\rm C}$ is parameterized as described in Ref. \citenum{Helbig_1DLDA}, whose parameterization was obtained using knowledge about the exact asymptomatic behavior of $\varepsilon_c(\rho)$ and fits to numerically exact quantum Monte Carlo calculations. Defining $r_s=1/(2\rho)$, the correlation contribution reads
 \begin{equation}\label{eq:CorrDens}
\varepsilon_C(r_s) = -\frac{1}{2} \frac{\ln(1 + \alpha r_s + \beta r_s^m)\left(r_s + E r_s^2\right)}{A + B r_s + C r_s^2 + D r_s^3},
\end{equation}
with the parameters shown in table \ref{tab:parameters}.
\begin{table}[htbp]
    \centering
    \caption{Parameters for the expression of the correlation energy density per particle, Eq.~\eqref{eq:CorrDens}. All parameters are taken from Ref. \citenum{Helbig_1DLDA}.}
    \label{tab:parameters}
    \begin{tabular}{lcccccccc}
        \hline
        $A$ & $B$ & $C$ & $D$ & $E$ & $\alpha$ & $\beta$ & $m$ \\
        \hline
        18.40 & 0.0 & 7.501 & 0.10185 & 0.012827 & 1.511 & 0.258 & 4.424 \\
        \hline
    \end{tabular}
\end{table}
 
The XC potential then reads
\begin{align}
    v_{\rm {XC}}^{\mathrm {LDA}}(x) &= \frac{\delta E^{\mathrm {LDA}}_{\rm {XC}}}{\delta \rho(x)} \nonumber \\
    \label{eq:LDA_functional_form}
    &=\varepsilon_{\rm {XC}}(\rho (x)) + \rho(x)\frac{\partial \varepsilon_{\rm {XC}}(\rho(x))}{\partial \rho(x)}.
\end{align}
We will refer to simulations done with this XC functional simply as (TD)DFT calculations. 

\subsection{Quantities of interest}

To quantify the quality of the wave functions, we compare the dipole moment and the HHG spectrum to that of a grid calculation. To be more precise, we consider two main quantities of interest:
\begin{enumerate}
    \item the time-dependent electronic dipole moment
\begin{equation}
    \langle\mu(t)\rangle=\langle\Psi(t)|\hat\mu|\Psi(t)\rangle=-\int\rho(x,t)x\text{d}x,
\end{equation}
\item the HHG spectrum in the approximate velocity form
\begin{equation} \label{eq:HHG}
S(\omega)\propto \omega^2 \left|  \int_{0}^{t_f}     \langle\mu(t)\rangle e^{\text i\omega t} \text{d}t \right|^2,
\end{equation}
obtained using a discrete Fourier transform, where we also made use of a Hann window function.~\cite{Hanning}
\end{enumerate}

\section{Numerical considerations}\label{sec:Numerics}

\subsection{Evaluation of the Rothe error}\label{sec:Evaluation}

For ease of implementation, we evaluate the Rothe error $r_{i+1}(\boldsymbol{c}(\boldsymbol{\alpha}),\boldsymbol{\alpha})$ on a grid.
Gaussian quadrature with 239 points is used in the region $(-17,17)$ a.u., while the trapezoidal rule is employed in the regions $[a,-17]$ a.u.~and $[17,b]$ a.u.~with a uniform grid with spacing $\Delta x=0.4$ a.u. For field strength $E_0=0.0534$ a.u.~we use $-a=b=320$, and for $E_0=0.1068$ a.u.~we use $-a=b=450$, except for the \LiHtwo\ TDDFT simulation where $-a=b=500$ is used (to avoid reflections). To compute the coefficient matrix $\boldsymbol{c}(\boldsymbol{\alpha})$ in a numerically stable way, we use {the SVD as described in Sec.~\ref{sec:VarPro}.} 
{While using an underlying grid to evaluate the Rothe error makes the implementation much simpler, it is a major bottleneck in terms of computational speed - $M(t)$ exponential functions need to be evaluated on each grid point, and the two-dimensional, dense Fock matrix needs to be constructed on a grid, for $\sim\!100$s of iterations per time point (see Sec.~\ref{sec:Propagation}). The grids needed for HHG calculations are very large, and the grid needs to be quite fine grained in order to capture all relevant oscillations. For that reason, with large number of Gaussians, the calculations become very slow, which restricts how many Gaussians can be used and how strong the fields can be. However, it is important to underline that for realistic calculations, this is not a weakness of Rothe's method with Gaussians in general, as grid size and resolution are normally not factors to consider. All matrix elements can be calculated analytically.\cite{MEST} Hence, this study should be understood as a proof-of-principle study.

With analytical integrals, Rothe's method scales quadratically in the number of Gaussians $M$. When uncorrelated Gaussians are used, the cost of calculating those integrals is similar in 3D as it is in 1D. While more Gaussians are needed for higher-dimensional systems, this is still a substantial advantage over grids, where the cost increases exponentially. A bottleneck of Rothe's method as opposed to other methods, though, is the fact that matrix elements of the squared Fock operator need to be calculated -- this means also that three-body integrals need to be calculated.  }
\subsection{Masking functions}\label{sec:MaskingFunctions}

We use a masking function {(or simply mask)} to absorb the outgoing wave function. {
The effect of a masking function $M(x)$ is to absorb the outgoing parts of the orbitals, i.e., after each time step the orbitals $\varphi_j(x,t)$ obtained from Rothe's method are replaced by
\begin{equation}\label{eq:Mask}
    \varphi_j(x,t) \leftarrow M(x)\varphi_j(x,t).
\end{equation}
With $0\le M(x)\le 1$, the propagation is effectively non‑unitary (non‑Hermitian) and the orbitals are no longer normalized. Indeed, masking functions are related to complex absorbing potentials (CAPs): to first order in the time step $\Delta t$, there is a one‑to‑one mapping between a masking function and an equivalent CAP.\cite{Giovannini}
}
The functional form {we use, following Ref.~\citenum{Sato_TDCAS_1D},} is:
\begin{equation}
M(x) = 
\begin{cases} 
\cos^{1/4}(\frac{\pi}{2}\frac{a_{\rm start}-x}{a_{\rm min}-a_{\rm start}}) & \text{for } a_{\rm min} < x < a_{\rm start}, \\
1 & \text{for } a_{\rm start} < x < b_{\rm start}, \\
\cos^{1/4}(\frac{\pi}{2}\frac{b_{\rm start}-x}{b_{\rm max}-b_{\rm start}}) & \text{for } b_{\rm start} < x < b_{\rm max}, \\
0 & \text{otherwise}.
\end{cases}
\end{equation}
where we choose the ``start'' of the mask to be at $a_{\rm start}=0.85\times a_{\rm min}$ and $b_{\rm start}=0.85\times b_{\rm max}$. Here, $ b_{\rm max}=b-5$ and $a_{\rm min}=a+5$ to take into account that Gaussians might slightly extend beyond $b_{\rm max}/a_{\rm min}$. 
As we work with Gaussians, direct application of a mask is not feasible. Instead, we follow the approach described in Ref.~\citenum{schrader2025HeHe}. Directly multiplying Gaussians by a mask on the coordinate grid is straightforward, but the masked orbitals are no longer strictly linear combinations of Gaussians. Hence, we refit each masked orbital by a new linear combination of Gaussians by minimizing the $L^2$ norm between the orbitals expressed as Gaussians and the non-Gaussian masked orbitals. This is done in the same way as for the determination of the initial orbitals, using the unmasked orbitals as a starting guess. 

\subsection{Norm and orthogonality conservation}\label{sec:conservationrules}

Rothe's method approximates Crank-Nicolson propagation, but unless the Rothe error is exactly 0, conservation of norm (and energy, for a time-independent Hamiltonian or Fock operator) as well as orbital orthogonality is not guaranteed. To overcome this problem, we re-orthonormalize the orbitals at each time step by updating the linear coefficients using Löwdin symmetric orthogonalization,~\cite{Mayer_Lowdin} followed by a renormalization of the orbitals. Löwdin symmetric orthogonalization ensures that the orbitals change as little as possible in a least-square sense. We have observed that the deviation of the optimized orbitals from both normality and orthogonality after a single time step is very small, hence this procedure only leads to a very minor increase in the Rothe error.
Orthonormalization is carried out before and after application of the masking function. The renormalization is with respect to the norm of the orbitals at the previous time point and after the mask was applied, respectively. I.e., the loss of unit norm due to the masking function is respected.

\subsection{Propagation}\label{sec:Propagation}

We use a time step $\Delta t=0.05$ in all simulations, with the initial state represented by the ground-state Gaussians and $2$ Gaussians with $a=\sqrt{1/2},b=p=0$ placed on each side of the molecule. The $4$ additional Gaussians are placed at $\mu\in\{-7,-5,5,7\}$ and at $\mu\in\{-15,-13,13,15\}$ for LiH and \LiHtwo, respectively. We kept the nonlinear coefficients of the initial state frozen at all times, as even minor changes in these parameters can lead to large changes in the Rothe error which, in turn, can lead to numerical issues for the optimization scheme.

As a starting guess for the nonlinear coefficients at time $t_{i+1}$, we use the nonlinear coefficients from the previous time point $t_i$, and except for the initial time point or when the number of Gaussians changed, we add a fraction of the change going from ${t_{i-1}}$ to $t_i$, i.e.,
\begin{equation}
\boldsymbol{\alpha}(t_{i+1})_{\text{init}}=\boldsymbol{\alpha}(t_i)+\delta(\boldsymbol{\alpha}(t_i)-\boldsymbol{\alpha}(t_{i-1})).
\end{equation}
The parameter $\delta$ is obtained using a line search. 
Similarly to our previous implementations of Rothe's method,~\cite{schrader2024time,schrader2025HeHe} we solve a modified optimization problem, where the parameters are not allowed to change arbitrarily. That is, we optimize a set of transformed parameters $(\boldsymbol{\alpha}^{i+1})'$,
\begin{equation}\label{eq:constr_opt}
(\boldsymbol{\alpha}_{i+1})'_j=\arcsin\left(2\cdot\frac{\boldsymbol{\alpha}(t_{i+1})-{\min}_j}{{\max}_j-{\min}_j}-1\right),
\end{equation}
where
\begin{align}
{\min}_j&=(\boldsymbol{\alpha}(t_{i+1})_{\text{init}})_j-s|(\boldsymbol{\alpha}(t_{i+1})_{\text{init}})_j|-q,\\
{\max}_j&=(\boldsymbol{\alpha}(t_{i+1})_{\text{init}})_j+s|(\boldsymbol{\alpha}(t_{i+1})_{\text{init}})_j|+q.
\end{align}
We have chosen $s=0.1$, and $q=0.05$ for the width parameters $a_m$, while we use $q=0.1$ for $b_m,\mu_m$ and $p_m$ for all calculations considered. If a Gaussian has been added within the last five time steps, we set $s=0.5$ in order for the Gaussians to be able to rearrange more broadly. Furthermore, due to the underlying grid, which gives an effective limit for how wide or narrow the Gaussians can realistically be, we require that the width parameters $a_m$ must lie in $[a_{\text{min}},a_{\text{max}}]$ where $a_{\text{min}}=0.1$ and $a_{\text{max}}=2$ for all Gaussians that do not represent the ground state, i.e., all Gaussians with time-dependent nonlinear coefficients. For the high-accuracy calculation of LiH using TDHF for the intensity $I_0=4\times 10^{14} \WcmSq$, we set $a_{\text{min}}=0.04$ at $t=219$ a.u., as no stable results were obtainable otherwise. However, we do not consider this a limitation of Rothe's method. Rather, it is a limitation of the underlying grid---such a restriction would not be necessary if analytical integrals were used and Gaussians could move fully freely. {Indeed, we have observed that numerical instability stemming from colliding Gaussians (see sec. \ref{sec:Overcompleteness} ) is due to the restriction of the factor $a_{\text{min}}$, which determines how wide the Gaussians can become. This restriction added because the grid to evaluate the Rothe error would otherwise have to become very large, as the Rothe error needs to be evaluated to very large precision, which requires that the Gaussians are numerically zero at the boundaries of the grid. It is hence an artifact of the grid implementation, more than it is a shortcoming of Rothe's method with Gaussians.}
For the optimization, we use the SciPy~\cite{scipy} implementation of the BFGS algorithm, with
\begin{equation} 
H_0^{-1}=\text{diag}\left(|{\nabla_{\boldsymbol\alpha}r_{i+1}\left(\boldsymbol{\alpha}(t_{i+1})_{\text{init}}\right)}|+\delta_{\text{num}}\right)^{-1},\end{equation}
as a guess for the inverse of the Hessian matrix at the first iteration, ensuring that the optimization is scale invariant. We set $\delta_{\text{num}}=10^{-14}$, which is sufficient to ensure invertibility. {We set the maximal number of iterations in the BFGS algorithm per time step to $300$, which we observe to be sufficient.}

A Gaussian is added if the Rothe error lies above a given threshold $\varepsilon_{\Delta t}/N_T$ after optimization. Similarly, Gaussians that barely contribute to the Rothe error are replaced by more contributing Gaussians. We describe how this is done in Sec.~\ref{sec:AddGaussians}. 

\subsection{Addition of Gaussians}\label{sec:AddGaussians}

The accuracy of the time propagation is monitored by the Rothe error $r_{i+1}$, Eq.~\eqref{eq:RotheError}. When this error exceeds the prescribed tolerance, $\varepsilon_{\Delta t}/N_T$, the size of the Gaussian basis is increased by one.
To do so, new candidate Gaussians are generated by sampling each nonlinear parameter ($a$, $b$, $p$, and $\mu$) from a distribution constructed from the current set of parameters via the Stochastic Variational Method.~\cite{Bubin_Adamowicz_ECG2013} That is, we generate $K$ candidate parameter sets $\{\boldsymbol{\alpha}_{M+1}^{(k)}\}_{k=1}^K$, where a candidate for a given parameter is drawn from a distribution proportional to
\[
p(\boldsymbol{\alpha}_{M+1})\propto\sum_{m=1}^{M}\exp\!\left[-\frac{(\boldsymbol{\alpha}_{M+1}-\boldsymbol{\alpha}_{m})^2}{2\boldsymbol{\alpha}_m^2}\right],
\]
where $\{\boldsymbol{\alpha}_{m}\}_{m=1}^M$ are the current parameter values and all expressions are to be interpreted element-wise. For each candidate set $\boldsymbol{\alpha}^{(k)}_{M+1}$, the effect of adding the corresponding Gaussian to the current basis is evaluated by computing the new Rothe error with the added Gaussian. Here, we used $K=500$. The candidate parameters that yield the lowest error are selected as a starting guess for the Gaussian to be added. Its nonlinear coefficients are then optimized using the BFGS algorithm (without constraints, i.e. not using Eq.~\eqref{eq:constr_opt}), 
which is followed by re-optimization of the nonlinear coefficients of all Gaussians following the re-parameterization in Eq.~\eqref{eq:constr_opt}. This addition is a numerically challenging process, and occasionally, no better solution is found than the initial parameter---in which case the unoptimized parameters are used. Even if that is the case, the added Gaussian then increases the variational space at the next time point, which leads to an overall decreasing Rothe error.

We also apply this addition procedure whenever a Gaussian barely contributes: That is, when the (optimized) Rothe error upon removal of one Gaussian increases by less than a factor of ${\kappa=}1.1$, that Gaussian is redundant. In that case, it is removed and replaced by a new Gaussian.

\subsection{Addressing overcompleteness and colliding Gaussians}\label{sec:Overcompleteness}

Two types of issues in the underlying basis cause numerical challenges. The first issue is that two Gaussians $g_k$ and $g_l$ can ``collide'' in the sense that they have almost the same nonlinear coefficients and, therefore, $S_{kl}=|\braket{g_k|g_l}| \approx 1$. The second issue is that the overlap matrix $\boldsymbol{S}$ may have very small eigenvalues. Both issues lead to numerical instability {which we observe by discontinuous jumps in expectation values $\langle A(t)\rangle$ and extremely large Rothe errors and large number of iterations required for convergence}. To overcome these issues, we implemented a scheme where at the beginning of each time step, using parameters $\boldsymbol{\alpha}(t_{i+1})_{\text{init}}$ giving rise to orbitals $\varphi_j$, we check whether there is an element $|S_{kl}|>s_{\text{max}}$ in the overlap matrix, or whether the lowest eigenvalue $\lambda_0$ of the overlap matrix is less than a threshold value $\lambda_{\text{min}}$.  
In each case, we remove a Gaussian from the nonlinear basis. If $\lambda_0<\lambda_{\text{min}}$, we take the Gaussian $g_k$ whose removal leads to the largest lowest eigenvalue $\lambda_0^{(k)}$. If $|S_{kl}|>s_{\text{max}}$, we take the Gaussian whose removal leads to a lesser change in the sum over the orbital errors $$\sum_{j=1}^N{\norm{\varphi_j-\varphi^{(k)}_j}}^2,$$ where $\varphi^{(k)}_j$ is the orbital that is as close as possible in a least-squares sense to orbital $\varphi_j$ when Gaussian $k$ is removed. We only remove unfrozen Gaussians, i.e., those not representing the ground state. We then re-optimize the basis with one Gaussian removed to fit to the orbitals, making sure that the Gaussians are sufficiently apart, i.e., we minimize
$$\epsilon(x)=\sum_{j=1}^N{\norm{\varphi_j-\tilde\varphi_j}}^2+\epsilon_p\left(\sum_{kl}P_{kl}(x)+\sum_k A_k(a_{\text{min}})\right),$$ with $x=0.95$ and
where $\tilde\varphi_j$ are the orbitals to be fitted. Here,
$$
P_{kl}(x) =
\begin{cases}
0 & \text{if } |S_{kl}| <x \\
 \frac{|S_{kl}|^2-x^2}{1-x^2}& \text{otherwise}\\
\end{cases}.
$$
and
$$
A_k(a_{\text{min.}})= 10\epsilon_p\left(\frac{a_{\text{min.}}}{|a_k|} - 1\right)^2,$$
where $\epsilon_p$ is the strength of the penalty term. 
The first penalty term $P_{kl}(x)$ makes sure that the new Gaussians are not near-linearly dependent,\cite{Bubin_Adamowicz_ECG2013} while the second penalty term $A_k(x)$ makes sure that the Gaussians do not acquire a too small width parameter. After that, we add one Gaussian to the basis by minimizing $\epsilon(0.7)$ over the parameters of just that Gaussian. Finally, we re-optimize $\epsilon(0.95)$ over all Gaussians again. If $\epsilon(0.95)>2\varepsilon_{\Delta t}/N_T$, this procedure is repeated, adding $2$ Gaussians instead of one (increasing the total number of Gaussians by one) as long as the total number of Gaussians is below $90$, and if the second Gaussian actually reduces the fitting error more than one does by a reduction of at least $10\%$---otherwise, only one is used. We then proceed using this new set of Gaussians as the new starting guess and do a Rothe propagation. This overall approach does increase the number of Gaussians, but this has the advantage that it leads to a larger number of functions available to represent each orbital, and hence lower Rothe errors. We have also observed that less linear dependency leads to fewer iterations in the minimization of the Rothe error. For the intensity $I_0=10^{14}\WcmSq$, we use $\epsilon_p=10^{-4}$ $s_{\text{max}}=0.99$ and $\lambda_{\text{min}}=10^{-10}$, for $I_0=4\times10^{14}\WcmSq$, we used $\epsilon_p=10^{-2}$, $s_{\text{max}}=0.99$ and $\lambda_{\text{min}}=10^{-9}$. 

In a few instances, especially for the high-accuracy calculations, we have observed that this procedure can lead to instabilities in expectation values, {often reflected in very large Rothe errors}. In those instances, we found that slightly perturbing some of the parameters in a trial-and-error fashion (e.g., using $\epsilon(0.9)$ instead of $\epsilon(0.95)$, reducing $\epsilon_p$ by orders of magnitude, or using $s_{\text{max}}=0.995$) can solve these issues.

The reason that this refitting is carried out before the Rothe optimization and not as part of the Rothe optimization itself, is that this scheme is less costly, as it only requires the overlap matrix and its derivatives. Furthermore, we have observed that penalizing the Rothe error in order to avoid near-overcompleteness as was done in Ref.~\citenum{schrader2024time} leads to numerical instability in some expectation values for too large values of $\epsilon_p$, and did not have any effect for too small values of $\epsilon_p$. Additionally, the Rothe error still captures all information about the deviation from exact propagation, i.e., there is no need to keep track of the fitting error.

{In figure \ref{fig:workflow}, we illustrate the workflow of Rothe's method for TDHF/TDDFT.
\begin{figure}
    \centering
    \includegraphics[width=0.8\linewidth]{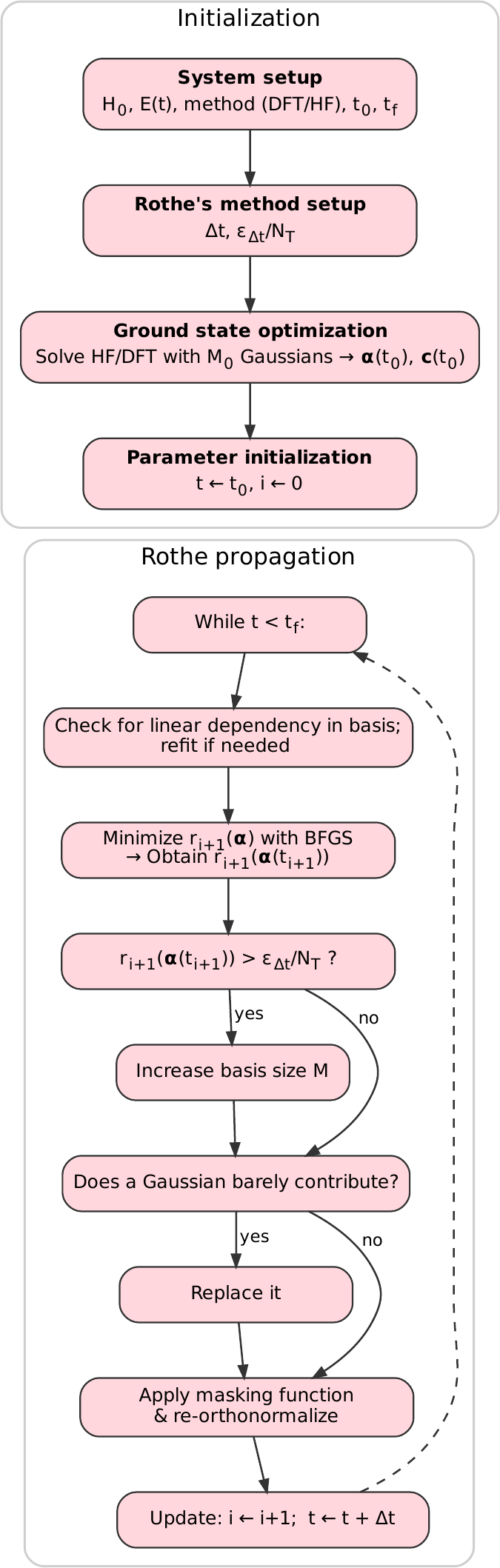}
    \caption{An illustration of the workflow for Rothe's method applied to the TDHF and TDDFT equations.}
    \label{fig:workflow}
\end{figure}
}
{\subsection{Hyperparameters, stability, and comparison to other methods}
In addition to the width restraints $a_{\mathrm{min.}}$ and $a_{\mathrm{max.}}$ that were already commented on, Rothe's method as implemented here relies on a set of hyperparameters, such as $\epsilon_p$, $s_{\mathrm{max.}}$, $\lambda_{\mathrm{min.}}$, $x=0.95$ and $\kappa=1.1$. This may seem like a lot of hyper parameters that need to be tuned, however, there is a general rationale behind how they were chosen. 

Once Gaussians have almost merged (i.e. their overlap is almost one), they don't move apart from each other anymore (because gradients are nearly the same), and $s_{\mathrm{max.}}$ should be chosen to allow Gaussians to get close, but not become (nearly) identical. the choice $s_{\mathrm{max.}}=0.99$ is a standard choice when constructing explicitly correlated Gaussian (ECG) basis sets for high-accuracy calculations \cite{Bubin_Adamowicz_ECG2013} and was also used by us in previous applications of Rothe's method.~\cite{schrader2024time} Other reasonable choices would be $s_{\mathrm{max.}}=0.98$ or $s_{\mathrm{max.}}=0.995$, i.e. a number close to $1$. 

The parameter $\epsilon_p$ is chosen to be similar in size to the per-time Rothe error $\varepsilon_{\Delta t}/N_T$. A possible choice would be $~0.01\varepsilon_{\Delta t}/N_T$. It should be chosen big enough to actually matter in relation to the Rothe error, but not too big for the optimization to become unstable.

The parameter $\lambda_{\mathrm{min.}}$ is a measure of global linear dependency in the basis, a too small value may lead to numerical instabilities related to large linear coefficients $\boldsymbol{c}$ arising in the VarPro procedure, and it is related to basis set overcompleteness for quantum chemistry calculations with Gaussian basis sets.  Choosing a too large would lead to an unnecessary refitting of the basis. Essentially, $\lambda_{\mathrm{min.}}=10^{-9}$ is safe choice, though we observed that we got stable results for $\lambda_{\mathrm{min.}}=10^{-10}$ for $I_0=10^{14}\WcmSq$.

The parameter $\kappa$ should be a number slightly larger than $1$ and no bigger than $\sim2$. If $\kappa$ is chosen to small, then the underlying basis might become bigger than necessary, as many barely-contributing Gaussians will remain. If $\kappa$ is chosen too large, then Gaussian replacement might be done a lot, increasing run time. $\kappa$ does in general not have a big impact on the results for a given $\varepsilon_{\Delta t}$.

The parameter $x=0.95$ should be less than $s_{\mathrm{max.}}$, but not too small. If chosen too small, then other Gaussians that have a large overlap might be affected too much and the basis would no longer be able to represent the orbitals, requiring more Gaussians, if chosen too small, then Gaussians might collide within few time steps, requiring further refitting. 

It should also be mentioned that there is a complicated interplay between the number of iterations necessary for convergence and the underlying basis. In general, we observe that stronger fields, more Gaussians, and more linear dependency in the basis leads to more iterations. Optimization becomes also more demanding when two Gaussians are nearly identical. The hyperparameters chosen here work towards alleviating this issue.

In contrast to regularization in methods such as vMCG, where regularization changes the underlying dynamics of the parameters and can lead to uncontrolled errors,\cite{feischl2024regularized} our regularization retains full error control as the Rothe error is a guide for how well the wave function at the next time step is represented. Instead of regularizing the equations that govern how the wave function at the next time step should be represented, we change the underlying basis to be able to carry out the optimization. The fact that we in a few cases had to perturb some of the parameters was precisely because the optimization did not manage to achieve a sufficiently small Rothe error otherwise. This is because the optimization process did not manage to reach a good minimum, and by slightly reforming the optimization landscape by changing some parameters, the optimization process managed to converge to a reasonable result. 
}

\subsection{Reference calculations}\label{sec:Grid}

For reference, we solve the TDHF (or TDDFT) equations of motion~\eqref{eq:TDHFDFT} on a discrete-variable-representation (DVR) grid.~\cite{lill1982discrete}
Specifically, we use sinc-DVR~\cite{colbert1992novel, boyd2001chebyshev, jones2016efficient} where $x \in [-L, L]$ is discretized uniformly as 
\begin{equation}
    x_{\alpha} = x_0+\alpha \Delta x, \, \alpha=0, \cdots, N_{\text{DVR}},
\end{equation}
where $x_0 \equiv -L$, $\Delta x = 2L/N_{\text{DVR}}$. Each grid point is associated with a sinc-DVR basis function given by~\cite{jones2016efficient}
\begin{equation}
    \xi_\alpha(x) \equiv \frac{1}{\sqrt{\Delta x}} \text{sinc} \left( \frac{x-x_\alpha}{\Delta x} \right),
\end{equation}
which satisfies the DVR property
\begin{equation}
    \label{eq:dvr_property}
    \xi_\alpha(x_\beta) = \frac{\delta_{\alpha \beta}}{\sqrt{\Delta x}},
\end{equation}
at the grid points.
Inner products are evaluated by quadrature,
\begin{equation}
    \label{eq:quad_rule}
    \braket{f|g} \equiv \int f^*(x) g(x) dx \approx \sum_{\alpha=0}^{N_{\text{DVR}}} f^*(x_\alpha) g(x_\alpha) w_\alpha,
\end{equation}
where the quadrature weights $w_\alpha=\Delta x$ for the sinc-DVR.
It follows from the DVR property \eqref{eq:dvr_property} that the sinc-DVR basis functions are orthonormal in the sense that  
\begin{equation}
    \braket{\xi_\alpha|\xi_\beta} = \delta_{\alpha \beta}.
\end{equation}

Matrix elements of the kinetic-energy operator in the sinc-DVR basis are given by~\cite{colbert1992novel, jones2016efficient}
\begin{equation}
    \label{eq:T_dvr}
    T_{\alpha\beta} \equiv \braket{\xi_\alpha|-\frac{1}{2}\frac{d^2}{dx^2}|\xi_\beta} =
    \begin{cases}
        \frac{\pi^2}{6 \Delta x^2} & \alpha=\beta, \\
        \frac{(-1)^{\alpha-\beta}}{\Delta x^2 (\alpha-\beta)^2} & \alpha \neq \beta,
    \end{cases}
\end{equation}
and the matrix elements of local potentials are diagonal, i.e.,
\begin{equation}
    \braket{\xi_\alpha|V(x)|\xi_\beta} = V(x_\alpha) \delta_{\alpha \beta}.
\end{equation}
Similarly, for the softened Coulomb potential defined in Eq.~\eqref{eq:soft_coulomb_def}
we have
\begin{align}
    \label{eq:u_abcd_dvr}
    \braket{\xi_\alpha \xi_\beta|W(x_1,x_2; Z, c)|\xi_\gamma \xi_\delta} &= W_{\alpha\beta} \delta_{\alpha \gamma} \delta_{\beta \delta},
\end{align}
where $W_{\alpha,\beta} \equiv W(x_\alpha, x_\beta; Z, c)$.

The time-dependent orbitals are expanded in the sinc-DVR basis as 
\begin{equation}
    \varphi_i(x,t) = \sum_{\alpha=0}^{N_{\text{DVR}}} C_{\alpha,i}(t) \xi_\alpha(x), \, i=1,\cdots,N_{\text{docc}},
\end{equation}
where $N_{\text{docc}} = N/2$ denotes the number of doubly occupied orbitals.
Inserting this expansion into Eq.~\eqref{eq:TDHFDFT} with $\hat{h}(\Phi(t),t)$ taken as the Fock operator,
multiplying by $\xi_\alpha(x)$, integrating over $x$ using the quadrature rule \eqref{eq:quad_rule}, and using Eq.~\eqref{eq:u_abcd_dvr}, the sinc-DVR TDHF equations become 
\begin{align}
    i \dot{C}_{\alpha, i}(t) &= \sum_{\beta} h_{\alpha \beta}(t) C_{\beta, i}(t) + \bar{W}^{\text{dir}}_{\alpha}(t;C(t)) C_{\alpha, i}(t)  \nonumber \\
    \label{eq:discrete_tdrhf_eq}
    &- \sum_{j=1}^{N_{\text{docc}}} \bar{W}^{\text{exc}}_{\alpha, j,i}(t;C(t))  C_{\alpha, j}(t).
\end{align}
Here, the direct and exchange mean-field matrix elements are given by
\begin{align}
    \bar{W}^{\text{dir}}_{\alpha}(t;C(t)) &\equiv 2\sum_{j=1}^{N_{\text{docc}}} \sum_{\beta} W_{\alpha, \beta} C^*_{\beta, j}(t) C_{\beta, j}(t) , \\
    \bar{W}^{\text{exc}}_{\alpha, j,i}(t;C(t)) &\equiv  \sum_{\beta} W_{\alpha, \beta} C^*_{\beta, j}(t) C_{\beta, i}(t),
\end{align}
and 
\begin{align}
    h_{\alpha \beta}(t) &= T_{\alpha \beta} + V(x_\alpha,t) \delta_{\alpha \beta}, \\
    V(x,t) &= -\mathlarger{\sum}_a^{N_a} W(x,X_a;Z_a,c) -E(t) x.
\end{align}
For TDDFT, the exchange term $\bar{W}^{\text{exc}}_{\alpha, j,i}(t;C(t))$ is replaced by the discrete XC potential given by Eq.~\eqref{eq:LDA_functional_form}, i.e.,
\begin{equation}
    \bar{W}^{\text{XC}}_\alpha(t; C(t)) = v^{\text{LDA}}_{\text{XC}}(x_\alpha; C(t)).
\end{equation}
Note that $\bar{W}^{\text{XC}}_\alpha(t; C(t))$ depends on $C(t)$ through the density, which in the DVR basis can be expressed as
\begin{equation}
    \rho(x,t) = \sum_{i=1}^{N_{\text{docc}}} |\phi_i(x,t)|^2 = \sum_i \sum_\alpha |C_{\alpha,i}(t)|^2.
\end{equation}

Defining the vector
\begin{equation}
    C(t) \equiv [C_{0,1} \cdots C_{N_{\text{DVR}},1} \cdots C_{0,N_{\text{docc}}} \cdots C_{N_{\text{DVR}},N_{\text{docc}}}]^T,
\end{equation}
we can recast Eq.~\eqref{eq:discrete_tdrhf_eq} in matrix form as 
\begin{equation}
    \dot{C}(t) = -iF(t;C(t))C(t) \label{eq:tdhf_dvr_matrix_form},
\end{equation}
where the matrix elements of the time-dependent HF and Kohn-Sham matrices are given by
\begin{align}
    F^{\text{HF}}_{(\alpha,i),(\beta,j)} &\equiv h_{\alpha \beta}(t) \delta_{i,j} + \bar{W}^{\text{dir}}_{\alpha}(t;C(t)) \delta_{\alpha \beta} \delta_{ij} \nonumber \\
    &- \bar{W}^{\text{exc}}_{\alpha, j,i}(t;C(t))\delta_{\alpha \beta}, \\
    F^{\text{KS}}_{(\alpha,i),(\beta,j)} &\equiv h_{\alpha \beta}(t) \delta_{i,j} \nonumber \\
    &+ \left(\bar{W}^{\text{dir}}_{\alpha}(t;C(t))  - \bar{W}^{\text{XC}}_{\alpha}(t;C(t)) \right)\delta_{\alpha \beta} \delta_{ij},
\end{align}
such that $(FC)_{\alpha,i} = \sum_{b,j} F_{(\alpha, i), (\beta,j)} C_{(\beta,j)}$ is given by the right-hand side of Eq.~\eqref{eq:discrete_tdrhf_eq}. 

To discretize Eq.~\eqref{eq:tdhf_dvr_matrix_form} in time, we use the {Crank--Nicolson} scheme described in Sec.~\ref{sec:RothesMethod}, resulting in each time step being a linear system
\begin{align}
    A(t_n, C^n) C^{n+1} = A^\dagger(t_n, C^n) C^n  \label{eq:tdrhf_CN_form},
\end{align}
where
\begin{align}
    A(t_n, \bar{C}) = I+\frac{i \Delta t}{2} F\left(t_n+\frac{\Delta t}{2}, \bar{C}\right).
\end{align}

Equation~\eqref{eq:tdrhf_CN_form} is solved iteratively with the biconjugate gradient stabilized (BiCGSTAB)~\cite{barrett1994templates}
method as implemented in the SciPy software library.~\cite{scipy}
As preconditioner for the BiCGSTAB method, we use the inverse of the field-free one-particle Hamiltonian
\begin{equation}
    M = \left(I+\frac{i \Delta t}{2} h(t_0)\right)^{-1},
\end{equation}
which is calculated once before the propagation.

We use the grid spacing $\Delta x=0.25$ a.u., the same grid extents as for the Rothe method (Sec.~\ref{sec:Evaluation}), and the same masking function (Sec.~\ref{sec:MaskingFunctions}). We also use the same time step, $\Delta t=0.05$ a.u. With these parameters and our choice of the mean-field, we expect the Rothe method to give almost identical results to the DVR calculation if sufficient Gaussians are used. Note that the DVR discretization is not strictly variational, which is reflected in the DFT results in Table~\ref{tab:energies}.

\section{Results}\label{sec:Results}

\subsection{Initial orbitals}\label{sec:InitOrbitals}

The initial ground state orbitals at $t=0$ were obtained using the method outlined in Sec.~\ref{sec:InitOrbs}. For LiH, $M=20$ Gaussians were used, for \LiHtwo, $M=34$ Gaussians were used. By the discussion in Sec.~\ref{sec:Propagation}, this means the propagation starts with $M=24$ Gaussians for LiH and with $M=38$ Gaussians for \LiHtwo.
The ground-state energies of the systems obtained using Gaussians are compared with the grid results in Table \ref{tab:energies}, where we also included the energies obtained on the DVR grid for $\Delta x=0.125$ a.u. to study the convergence on the grid. We see that the energy difference for different values of $\Delta x$ is comparable in magnitude to the difference between the grid calculations and the Gaussian calculations, with deviations below $10^{-6}$ Hartree, indicating that the Gaussian energies are very accurate.
\begin{table}[htbp]
    \centering
    \caption{Energies in atomic units for the LiH and \LiHtwo\ systems calculated using grid calculations and 20 (LiH) or 34 (\LiHtwo) Gaussians.}
    \label{tab:energies}
    \begin{tabular}{lccc}
        \hline
        \noalign{\vskip 0.1cm}
        & Grid ($\Delta x\!=\!0.25$) & Grid ($\Delta x\!=\!0.125$) & Gaussians \\
        \noalign{\vskip 0.1cm}\hline\noalign{\vskip 0.1cm}
        $E^{\text{HF}}_{\text{LiH}}$          & -7.0658152003  &-7.0658154283   & -7.0658154275 \\[0.05cm]
        $E^{\text{HF}}_{\text{(LiH)}_2}$        & -14.1372000890   & -14.1371996949    & -14.1371994904 \\[0.05cm]
        $E^{\text{DFT}}_{\text{LiH}}$         & -7.0506591074   & -7.0506594064    & -7.0506597247 \\[0.05cm]
        $E^{\text{DFT}}_{\text{(LiH)}_2}$       & -14.1162274678   &  -14.1162270155    & -14.1162267032   \\
        \noalign{\vskip 0.1cm}\hline
    \end{tabular}\end{table}

The field-free Rothe errors represent the inherent error due to the initial state not being an exact ground state of the effective Hamiltonian. They can be interpreted as an approximation to the sum of standard deviations for each orbital times the time step,~\cite{schrader2024time}
and provide us with an effective lower bound for the Rothe error in the presence of a field. As seen in Table \ref{tab:standard_deviations}, the field-free Rothe errors are
very small for the systems considered in this work.
\begin{table}[htbp]
    \centering
    \caption{Field-free Rothe error using a time step $\Delta t=0.05$.}
    \label{tab:standard_deviations}
    \begin{tabular}{lccccc}
        \hline
        \noalign{\vskip 0.1cm} 
        \text{Errors} & $r^{\text{HF}}_{\text{LiH}}$ & $r^{\text{HF}}_{\LiHtwo}$ & $r^{\text{DFT}}_{\text{LiH}}$ & $r^{\text{DFT}}_{\LiHtwo}$ \\
        \noalign{\vskip 0.1cm} \hline\noalign{\vskip 0.1cm} 
        \text{Values} & $1.34\times10^{-6}$ & $7.01\cdot10^{-6}$ & $4.88\cdot10^{-6}$ & $8.23\cdot10^{-6}$ \\
        \noalign{\vskip 0.1cm} \hline
    \end{tabular}
\end{table}

\subsection{HHG spectra, dipole moments and densities}

\begin{figure}[hbpt]
    \centering
    \includegraphics[width=\linewidth]{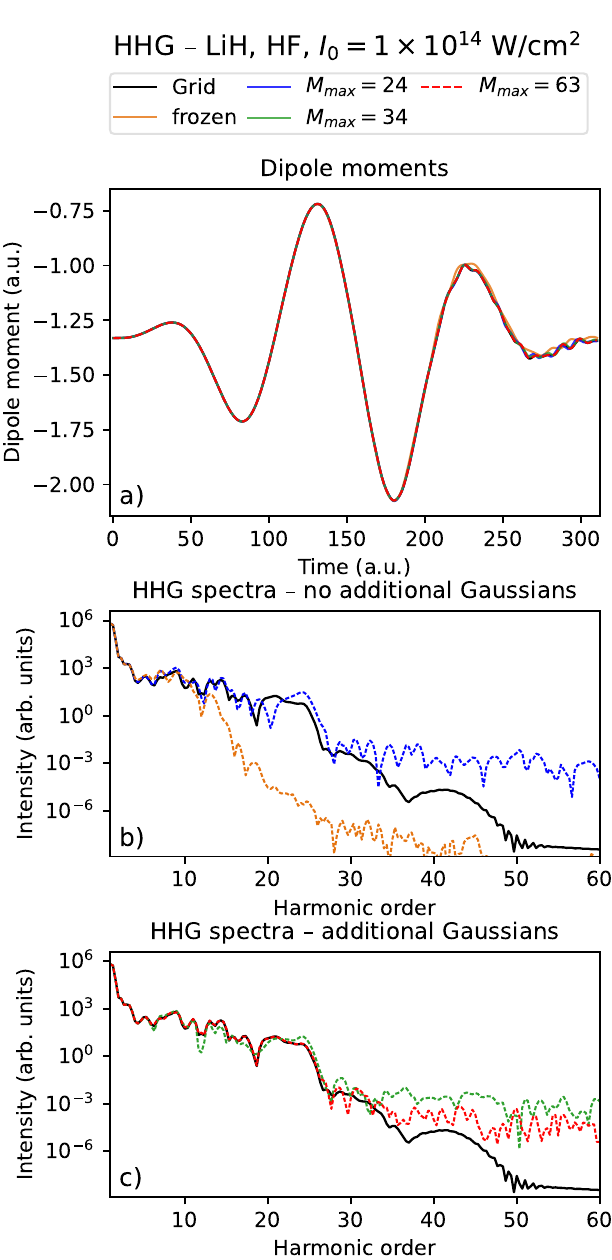}
    \caption{Time-dependent dipole moment (a) and HHG spectra (b, c) for the grid solution and Rothe's method with a varying number of Gaussians for LiH using TDHF with $I_0=10^{14}\WcmSq$.}
    \label{fig:HHG_LiH_HF_1}
\end{figure}

\begin{figure}[hbpt]
    \centering
    \includegraphics[width=\linewidth]{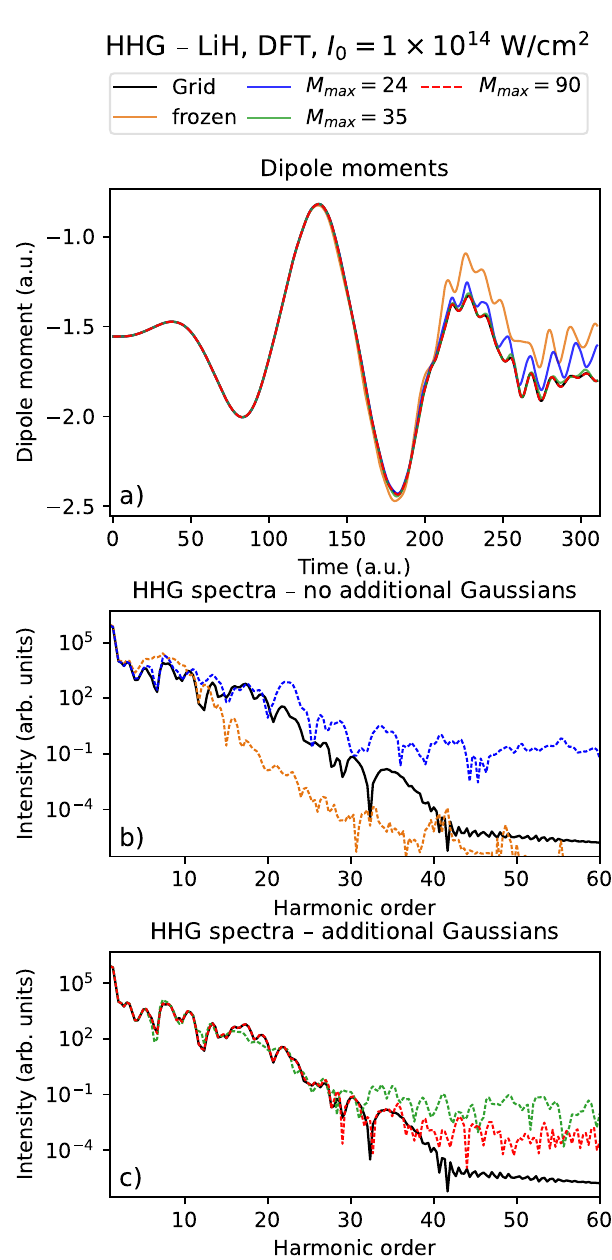}
    \caption{Time-dependent dipole moment (a) and HHG spectra (b, c) for the grid solution and Rothe's method with a varying number of Gaussians for LiH using TDDFT with $I_0=10^{14}\WcmSq$.}
    \label{fig:HHG_LiH_DFT_1}
\end{figure}

\begin{figure}[hbpt]
    \centering
    \includegraphics[width=\linewidth]{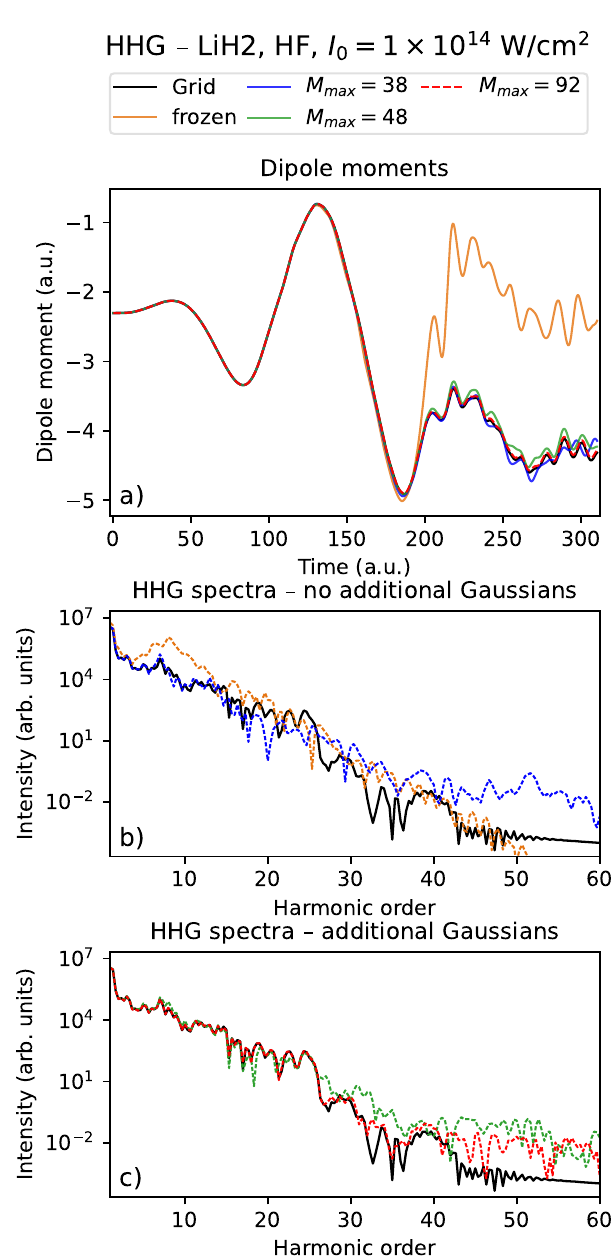}
    \caption{Time-dependent dipole moment (a) and HHG spectra (b, c) for the grid solution and Rothe's method with a varying number of Gaussians for \LiHtwo\ using TDHF with $I_0=10^{14}\WcmSq$.}
    \label{fig:HHG_LiH2_HF_1}
\end{figure}

\begin{figure}[hbpt]
    \centering
    \includegraphics[width=\linewidth]{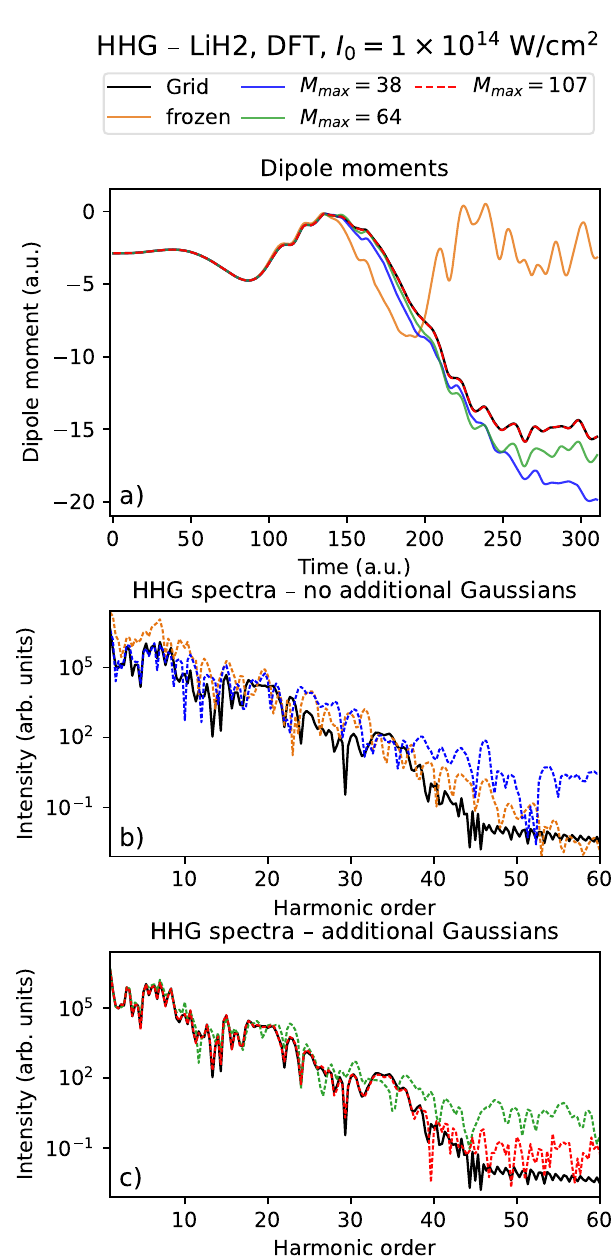}
    \caption{Time-dependent dipole moment (a) and HHG spectra (b, c) for the grid solution and Rothe's method with a varying number of Gaussians for \LiHtwo\ using TDDFT with $I_0=10^{14}\WcmSq$.}
    \label{fig:HHG_LiH2_DFT_1}
\end{figure}

We present results from different Rothe propagations defined in terms of the orbital parameters that are allowed to change:
\begin{enumerate}
    \item Only the linear coefficients are propagated, while all nonlinear parameters and the number of Gaussians are fixed at their initial values. This corresponds to frozen-Gaussian propagation.
    \item \label{type2} Only the $4$ additional Gaussians of the initial state and all linear coefficients are allowed to change, with no new Gaussians added during propagation.
        This corresponds to setting $\varepsilon_{\Delta t} = \infty$.
    \item As in point \ref{type2}, but with Gaussians added whenever the Rothe error is greater than $\varepsilon_{\Delta t}/N_T$.
\end{enumerate}
Table \ref{tab:Rothe_E0=1_summary} contains the information about the Rothe errors and the resulting final number of Gaussians, as well as the chosen threshold $\varepsilon_{\Delta t}$ for the intensity $I_0=10^{14}\WcmSq$. In addition, it also contains the difference of the density (Eq.~\eqref{eq:Density}) between the grid calculation  $\int|\Delta \rho| \mathrm dx= \int |\rho_{\mathrm{grid}}(x)-\rho_{\mathrm{gauss}}(x)| \mathrm dx$ at $t=t_f=310.25$ a.u..
\begin{table}[htbp]
    \centering
    \caption{Summary of resulting number of Gaussians and Rothe errors for different systems and methods using the intensity $I_0=10^{14}\WcmSq$. M$_{\text{max}}$ stands for the number of Gaussians at the final time $t_f$, $\varepsilon_{\Delta t}$ is the threshold for the addition of an additional Gaussian, and $r_{\text{tot.}}$ is the cumulative Rothe error. \textit{frozen?} indicates whether all nonlinear coefficients were kept  frozen. $\int|\Delta \rho| \mathrm dx= \int |\rho_{\mathrm{grid}}(x)-\rho_{\mathrm{gauss}}(x)| \mathrm dx$ is the cumulative difference between the grid density $\rho_{\mathrm{grid}}$ and the density calculated from Gaussians $\rho_{\mathrm{gauss}}$ at the final time $t=t_f=310.25$ a.u..}
    \label{tab:Rothe_E0=1_summary}
    \begin{tabular}{lcccccc}
        \hline
        \noalign{\vskip 0.1cm} 
        \text{System} & \text{Method} & M$_{\text{max}}$ & $\varepsilon_{\Delta t}$ & $r_{\text{tot.}}$ & frozen? & $\int|\Delta \rho| \mathrm dx$\\
        \noalign{\vskip 0.1cm} \hline\noalign{\vskip 0.1cm} 
        \text{LiH} & \text{HF} & 24 & -- & 0.27 &yes & 7.243$\cdot 10^{-3}$\\
        \text{LiH} & \text{HF} & 24 & $\infty$ & 0.28 &no &  6.841$\cdot 10^{-3}$\\
        \text{LiH} & \text{HF} & 34 & 0.5 & 0.091 &no & 2.045$\cdot 10^{-3}$\\
        \text{LiH} & \text{HF} & 63 & 0.1 & 0.027 &no & 2.681$\cdot 10^{-4}$\\

        \text{LiH} & \text{DFT} & 24 & -- & 1.1&yes &5.459$\cdot 10^{-2}$\\
        \text{LiH} & \text{DFT} & 24 & $\infty$ & 0.95&no& 3.962$\cdot 10^{-2}$\\
        \text{LiH} & \text{DFT} & 35 & 1 & 0.36&no& 6.948$\cdot 10^{-3}$\\
        \text{LiH} & \text{DFT} & 90 & 0.2 & 0.09&no& 3.903$\cdot 10^{-4}$\\
        
        \text{\LiHtwo} & \text{HF} & 38 & -- & 5.0&yes& 1.850$\cdot 10^{-1}$\\
        \text{\LiHtwo} & \text{HF} & 38 & $\infty$ & 1.3&no& 8.636$\cdot 10^{-2}$\\
        \text{\LiHtwo} & \text{HF} & 48 & 3.0 & 0.90&no& 5.451$\cdot 10^{-2}$\\
        \text{\LiHtwo} & \text{HF} & 92 & 0.6 & 0.15&no& 2.006$\cdot 10^{-3}$\\

        \text{\LiHtwo} & \text{DFT} & 38 & -- & 20.4&yes& 7.396$\cdot 10^{-1}$\\
        \text{\LiHtwo} & \text{DFT} & 38 & $\infty$ & 9.6&no& 7.686$\cdot 10^{-1}$\\
        \text{\LiHtwo} & \text{DFT} & 64 & 10 & 3.0&no& 5.212$\cdot 10^{-1}$\\
        \text{\LiHtwo} & \text{DFT} & 107 & 2 & 0.77&no& 1.414$\cdot 10^{-2}$\\
        \noalign{\vskip 0.1cm} \hline
    \end{tabular}
\end{table}
Figures \ref{fig:HHG_LiH_HF_1} and \ref{fig:HHG_LiH_DFT_1}  show the time-dependent dipole moments and HHG spectra using Rothe's method for LiH with an intensity $I_0=10^{14}\WcmSq$ using a varying number of Gaussians for TDHF and TDDFT, respectively, compared with the grid reference simulations. Figures \ref{fig:HHG_LiH2_HF_1} and \ref{fig:HHG_LiH2_DFT_1} show the corresponding data for the dimer \LiHtwo. 

In all four figures, one sees that the grid dipole moments and HHG spectra are better reproduced by increasing the number of Gaussians. In particular, we see that with the maximal number of Gaussians considered (the data shown using red graphs in all figures), the dipole moments follow the grid calculation closely, and the HHG spectra agree very well with the grid up to approximately the $30$th to $40$th harmonic. We observe that the results using the frozen Gaussians are qualitatively wrong for all systems considered. However, just using 4 thawed Gaussians instead leads to a large improvement of the dipole moments, which follow the general shape of the grid solution. The HHG spectra are quantitatively correct up to the $\sim\!15$th harmonic. For all systems considered, we see that there is some noise in the very high harmonics. The spectra no longer decrease and shows noisy behavior from approximately the 30th to 40th harmonic for all systems and number of Gaussians considered. However, we see that adaptively increasing the number of Gaussians both reduces the noise and pushes the number of correct harmonics higher, especially for \LiHtwo. From table \ref{tab:Rothe_E0=1_summary}, we observe that the density obtained using Gaussians gets very close to the grid density as the number of Gaussians is increased, with the error decreasing by up to two orders of magnitude compared to the frozen Gaussian calculation. We observe that the densities with 4 thawed Gaussians are only marginally better than just keeping all Gaussians frozen --- indicating that a big improvement in expectation values and HHG spectra is obtained even when the resulting densities and orbitals are still quite different from the converged solution.

\begin{figure}[hbpt]
    \centering
    \includegraphics[width=\linewidth]{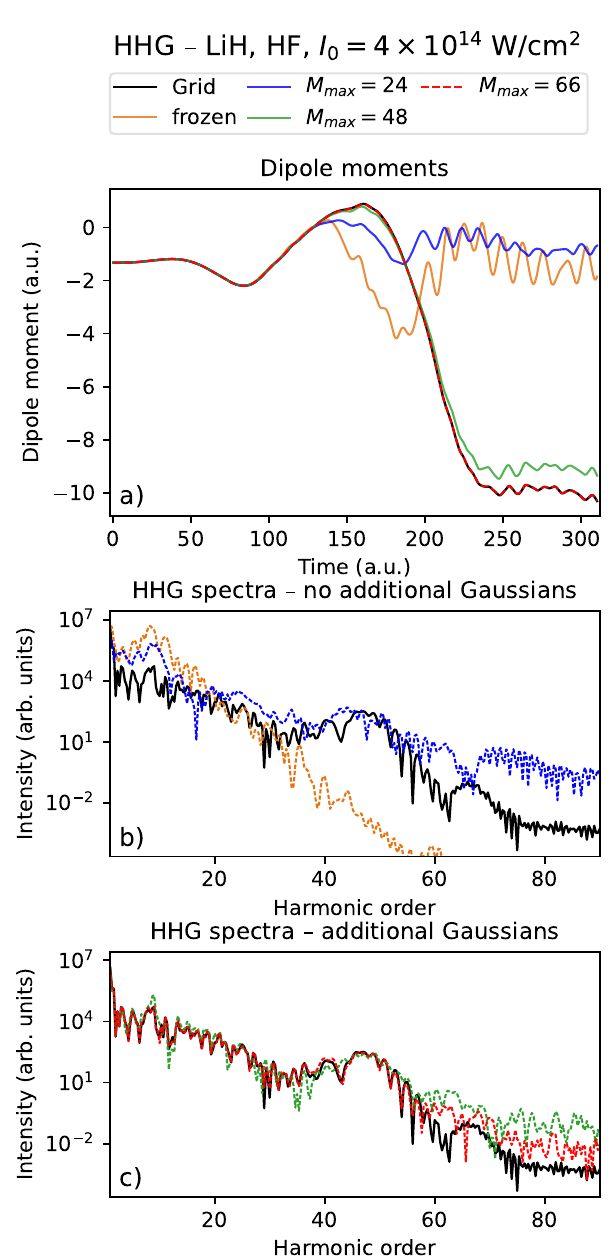}
    \caption{Time-dependent dipole moment (a) and HHG spectra (b, c) for the grid solution and Rothe's method with a varying number of Gaussians for LiH using TDHF with intensity $I_0=4\times10^{14}\WcmSq$.}
    \label{fig:HHG_LiH_HF_4}
\end{figure}

\begin{figure}[hbpt]
    \centering
    \includegraphics[width=\linewidth]{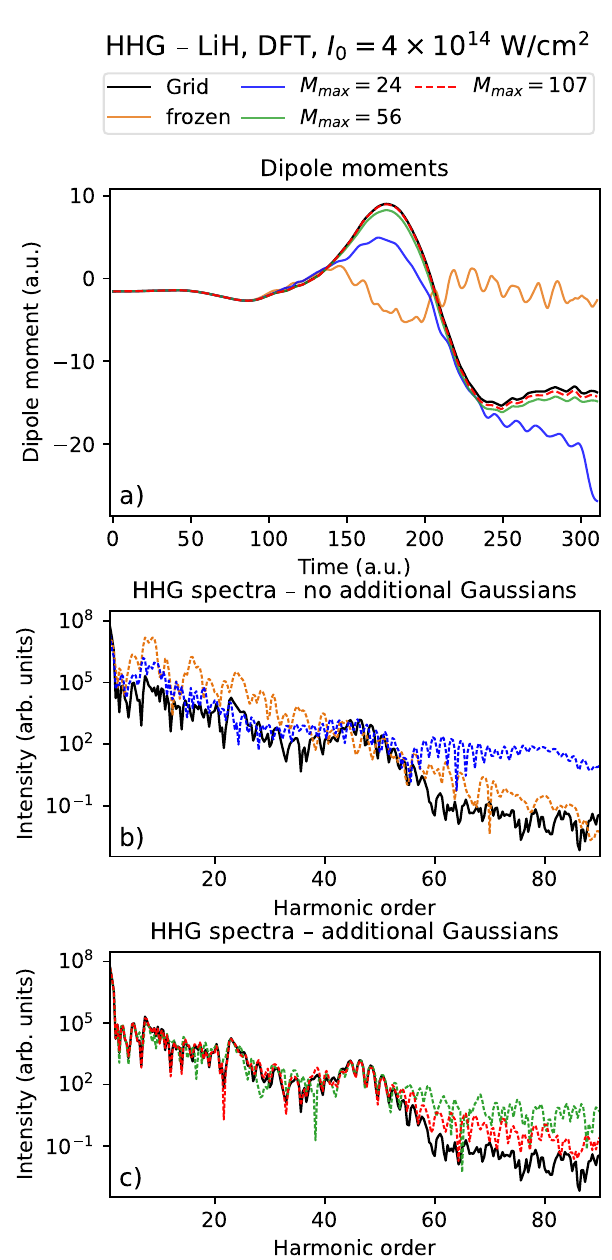}
    \caption{Time-dependent dipole moment (a) and HHG spectra (b, c) for the grid solution and Rothe's method with a varying number of Gaussians for LiH using TDDFT with intensity $I_0=4\times10^{14}\WcmSq$.}
    \label{fig:HHG_LiH_DFT_4}
\end{figure}

\begin{figure}[hbpt]
    \centering
    \includegraphics[width=\linewidth]{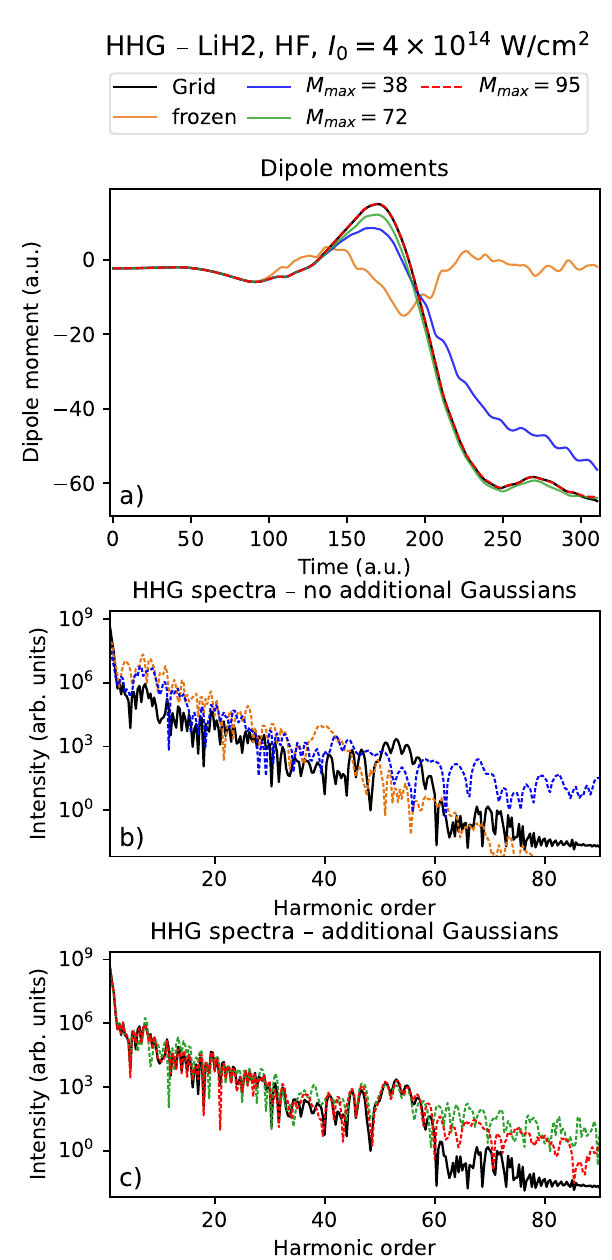}
    \caption{Time-dependent dipole moment (a) and HHG spectra (b, c) for the grid solution and Rothe's method with a varying number of Gaussians for \LiHtwo\ using TDHF with intensity $I_0=4\times10^{14}\WcmSq$.}
    \label{fig:HHG_LiH2_HF_4}
\end{figure}

\begin{figure}[hbpt]
    \centering
    \includegraphics[width=\linewidth]{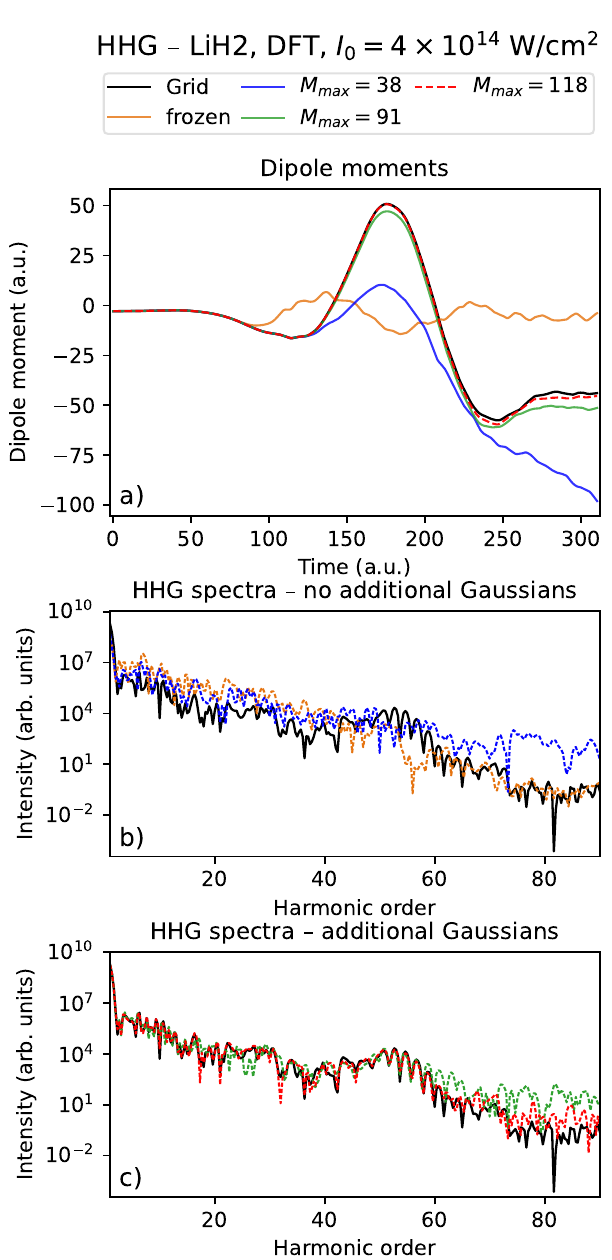}
    \caption{Time-dependent dipole moment (a) and HHG spectra (b, c) for the grid solution and Rothe's method with a varying number of Gaussians for \LiHtwo\ using TDDFT with intensity $I_0=4\times10^{14}\WcmSq$.}
    \label{fig:HHG_LiH2_DFT_4}
\end{figure}

Table \ref{tab:Rothe_E0=4_summary} and \Cref{fig:HHG_LiH_HF_4,fig:HHG_LiH_DFT_4,fig:HHG_LiH2_HF_4,fig:HHG_LiH2_DFT_4} show the corresponding results for the intensity $I_0=4\times10^{14}\WcmSq$. Compared to the intensity $I_0=10^{14}\WcmSq$, we see that with the highest number of Gaussians considered, there is almost quantitative agreement with the grid calculations for the dipole moments for LiH \LiHtwo\ using TDHF, while they appear close to converged for TDDFT. There is a clear improvement in the dipole moment data with when the number of Gaussians is increased. Except for LiH using TDHF, we observe that using just 4 thawed Gaussians are sufficient to at least qualitatively reproduce the general shape of the dipole moment, though to a much lesser degree than for the weaker field. Similarly, we observe for the HHG data that we are able to obtain HHG spectra that quantitatively agree with the grid data up to the 60th harmonic using the highest number of Gaussians considered, with improving results in both the positions of the peaks and the tail as the number of Gaussians increases. Just as for $I_0=10^{14}\WcmSq$, we observe that the difference to the grid density decreases by up to two orders of magnitude as the number of Gaussians is increased to the maximal value. For \LiHtwo\, however, we observe that the densities have not quite converged, even though good agreement in dipole moments and HHG spectra is obtained for TDHF.  

\begin{figure*}[htbp]
  \centering
  \includegraphics[width=\textwidth]{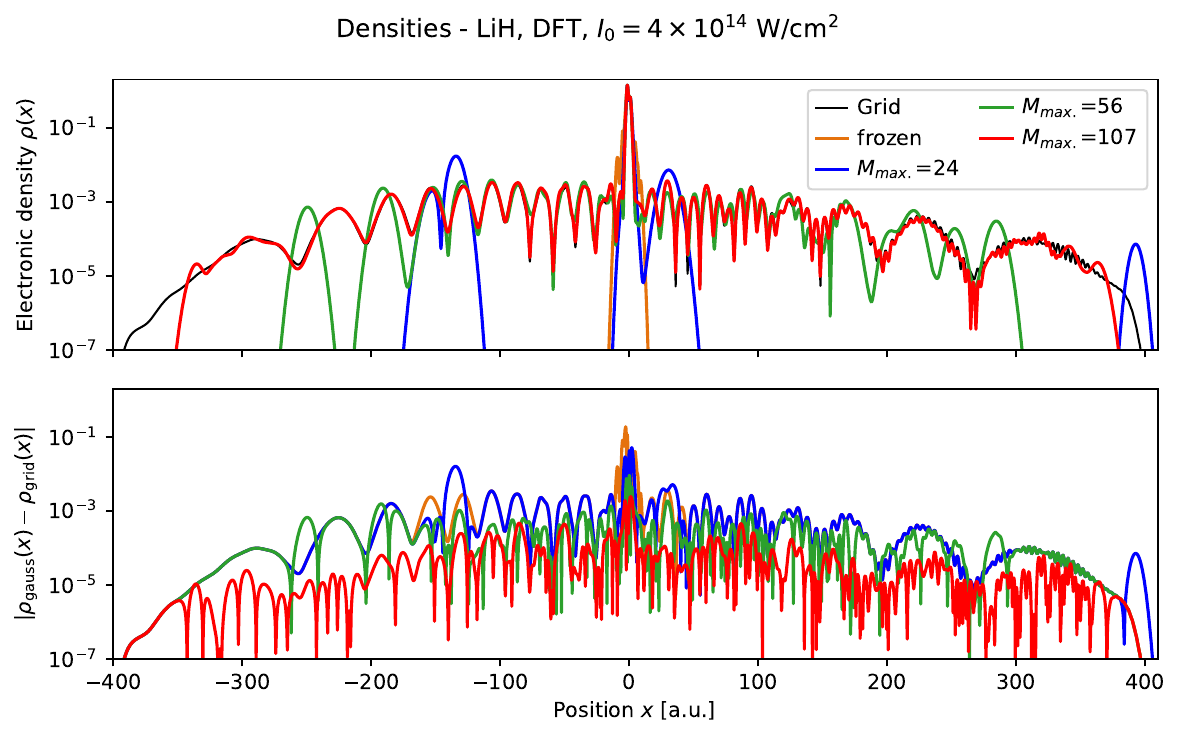}
  \caption{Electronic density $\rho(x)$ for varying number of Gaussians obtained using Rothe's method compared to the grid solution using TDDFT for LiH with intensity $I_0=4\times10^{14}\WcmSq$ at time $t=t_f=310.25$ a.u.. Top: Electronic densities, Bottom: The difference between the Gaussian electronic densities and the grid electronic densities. }
  \label{fig:DFT_LiH_4_density}
\end{figure*}

Finally, in figure \ref{fig:DFT_LiH_4_density}, we show the electron density $\rho(x)$ at the final time $t=t_f=310.25$ a.u. for LiH obtained using TDDFT with intensity $I_0=4\times10^{14}\WcmSq$, as well as the difference to the grid calculation. We observe that as the number of Gaussians is increased, finer details of the electronic density are reproduced, as can be seen with the oscillations for $x\in[100,250]$ --- using $M_{\mathrm{max.}}=56$ Gaussians, those are not reproduced, but using $M_{\mathrm{max.}}=107$ Gaussians, they are. Furthermore, we see that the error in the density reduces not only on average as shown in table \ref{tab:Rothe_E0=4_summary} as the number of Gaussians is increased, but it reduces everywhere --- showing that by minimizing the Rothe error, that convergence towards the exact wave function happens over all of space, not just in a specific region.

\begin{table}[htbp]
    \centering
    \caption{Summary of resulting number of Gaussians and Rothe errors for different systems and methods using the intensity $I_0=4\times10^{14}\WcmSq$. M$_{\text{max}}$ stands for the number of Gaussians at the final time $t_f$, $\varepsilon_{\Delta t}$ is the threshold for the addition of an additional Gaussian, and $r_{\text{tot.}}$ is the cumulative Rothe error. \textit{frozen?} indicates whether all nonlinear coefficients were kept  frozen. $\int|\Delta \rho| \mathrm dx= \int |\rho_{\mathrm{grid}}(x)-\rho_{\mathrm{gauss}}(x)| \mathrm dx$ is the cumulative difference between the grid density $\rho_{\mathrm{grid}}$ and the density calculated from Gaussians $\rho_{\mathrm{gauss}}$.}
    \label{tab:Rothe_E0=4_summary}
    \begin{tabular}{lcccccc}
        \hline
        \noalign{\vskip 0.1cm} 
        \text{System} & \text{Method} & M$_{\text{max}}$ & $\varepsilon_{\Delta t}$ & $r_{\text{tot.}}$& frozen? &$\int|\Delta \rho| \mathrm dx$\\
        \noalign{\vskip 0.1cm} \hline\noalign{\vskip 0.1cm} 
        \text{LiH} & \text{HF} & 24 & -- & 7.8&yes &4.671$\cdot 10^{-1}$\\
        \text{LiH} & \text{HF} & 24 & $\infty$ & 5.9&no&3.196$\cdot 10^{-1}$ \\
        \text{LiH} & \text{HF} & 48 & 5 & 1.1 &no&1.078$\cdot 10^{-1}$\\
        \text{LiH} & \text{HF} & 66 & 0.8 & 0.22 &no&4.766$\cdot 10^{-3}$\\

         \text{LiH} & \text{DFT} & 24 & -- & 21.6&yes& 1.080\\
        \text{LiH} & \text{DFT} & 24 & $\infty$ & 7.3&no&7.652$\cdot 10^{-1}$ \\
        \text{LiH} & \text{DFT} & 56 & 10 & 2.7&no& 2.055$\cdot 10^{-1}$\\
        \text{LiH} & \text{DFT} & 107 & 3 & 0.82 &no&4.712$\cdot 10^{-2}$\\

        \text{\LiHtwo} & \text{HF} & 38 & -- & 36.5&yes&2.138 \\
        \text{\LiHtwo} & \text{HF} & 38 & $\infty$ & 16.3 &no&1.974\\
        \text{\LiHtwo} & \text{HF} & 72 & 30 & 6.9 &no&1.031\\
        \text{\LiHtwo} & \text{HF} & 95 & 15 & 2.0 &no&1.459$\cdot 10^{-1}$\\

        \text{\LiHtwo} & \text{DFT} & 38 & -- & 57.9&yes&3.651 \\
        \text{\LiHtwo} & \text{DFT} & 38 & $\infty$ & 35.4&no&3.350 \\
        \text{\LiHtwo} & \text{DFT} & 91 & 30 & 8.1 &no&1.665\\
        \text{\LiHtwo} & \text{DFT} & 118 & 10 & 3.4 &no&5.509$\cdot 10^{-1}$\\
        \noalign{\vskip 0.1cm} \hline
    \end{tabular}
\end{table}

\section{Discussion}\label{sec:Discussion}

We observe that the quality of the HHG spectra improves with an increasing number of Gaussians. In particular, we observe for the intensity $I_0=10^{14}\WcmSq$ that just 4 additional, thawed Gaussians strongly improve the quality of the spectrum and allows the dipole moment to take a qualitatively correct shape. Further increasing the number of Gaussians gives quantitative agreement with the grid calculation. While our calculations do not reproduce grid results exactly, this shows that just allowing few thawed Gaussians can strongly improve the quality of a calculation where ionization needs to be considered -- and further increasing this number will lead to strongly improved results that are comparable to grid quality. This aligns with our observations from Rothe calculations for the hydrogen atom in a strong field.~\cite{schrader2024time} We observe that the TDDFT calculations give rise to more complicated dynamics -- this can be seen both from the magnitude of the change in the dipole moment for the intensity $I_0=10^{14}\WcmSq$, as well as from the fact that a similar number of basis functions gives rise to a much larger Rothe error, as can be seen both for the initial state and the cumulative Rothe error as well as the observation that the difference between the Gaussian densities compared to the grid density is much larger for the TDDFT calculations (see \cref{tab:standard_deviations,tab:Rothe_E0=1_summary,tab:Rothe_E0=4_summary}). 

We make the same observation comparing \LiHtwo\ to LiH. Nevertheless, we observe that approximately $\sim\! 50$ additional Gaussians at the final time are sufficient to give quantitatively correct HHG spectra up to the thirtieth harmonic. This indicates that it is not so much the value of the cumulative Rothe error that indicates the quality of the results, but that one instead should check that observables converge with an increasing number of Gaussians and decreasing Rothe errors. Indeed, we observe that the cumulative Rothe error is orders of magnitude larger than the difference between the densities, which is a better way to measure the convergence towards the exact solution when such a reference is available. Nevertheless, we also see that as the number of Gaussians is increased, that the relative improvement of the Rothe error is similar in magnitude to the improvement in the density difference - indicating that an improvement in the Rothe error by a given factor roughly represents the same improvement in the electron density. {While we usually observe that the cumulative Rothe error $r_{\text{tot.}}$ decreases as the basis becomes more flexible, this is not guaranteed. For LiH using TDHF at $I_0=10^{14}$\,\WcmSq\ (see table \ref{tab:Rothe_E0=1_summary}), $r_{\text{tot.}}$ increases when four Gaussians are thawed. For \LiHtwo\ using TDDFT at $I_0=10^{14}$\,\WcmSq, $r_{\text{tot.}}$ is roughly halved when $4$ Gaussians are flexible, yet the final density error increases. These cases show that $r_{\text{tot.}}$ is an aggregate of the local errors $r_{i}(\boldsymbol{\alpha}(t_i),\boldsymbol{c}(t_i))$ over the trajectory, controlled by the step tolerance $\varepsilon_{\Delta t}$ (which in these examples is infinite). It is not a bound on the global wavefunction/orbital error or any single observable at $t_f$. A more flexible basis can keep $r_{i}(\boldsymbol{\alpha}(t_i),\boldsymbol{c}(t_i))$ small early but incur larger $r_{i}$ later in the pulse, which might reduce the quality of the final density, even if the overall $r_{\text{tot.}}$ looks better. Once sufficient adaptivity is enforced by a small enough threshold $\varepsilon_{\Delta t}$, $r_{\text{tot.}}$ and the discrepancy in the final density improve together (see table \ref{tab:Rothe_E0=1_summary} and fig. \ref{fig:DFT_LiH_4_density}). We would also like to point out that this behavior is not specific to Rothe’s method or to Gaussians; any basis that is insufficient at relevant times can show the same effect (in particular, the wave function at $t_f$ can be better in the strictly smaller basis).
}

{Our results indicate that even for more complicated systems, that the number of additional Gaussians necessary to accurately represent the dynamics seems to be manageable---for LiH and \LiHtwo\ using TDHF, 43 and 58 additional Gaussians, respectively, give the same qualitative agreement with the grid in the HHG spectrum.} The fact that the TDDFT calculations seem to require a similar number of additional Gaussians to obtain the same quality in the HHG spectra, which we observe e.g. using 34 (TDHF) or 35 (TDDFT) Gaussians for LiH using $I_0=10^{14}\WcmSq$, indicates that the increase in the basis set size for correlated methods should be manageable. Those observations are also true for the stronger field with intensity $I_0=4\times 10^{14}\WcmSq$. For the stronger field, however, we observe that far more Gaussians are needed to obtain qualitatively correct results---4 Gaussians are not sufficient, but $~30$ Gaussians give qualitatively correct behavior in the HHG spectrum and the dipole moment, though even more Gaussians are necessary to get quantitatively correct results. In general, we observe a systematic convergence of both the dipole moment and the HHG spectra for all systems, methods and field strengths considered when the number of Gaussians is increased.

{Comparing the convergence behavior to the 3D single-electron case,\cite{schrader2024time} we observe that the number of Gaussians is quite comparable. Indeed, while the 3D case required more Gaussians for comparable field strengths, it is not orders of magnitude more. Even though those are different models, this is promising in terms of taking Rothe's method for many-body systems to realistic 3D systems, indicating that the number of required Gaussians does not scale exponentially in the dimension, especially for symmetric systems. }
{As written in Sec.~\ref{sec:Evaluation}, due to the high computational cost of using a grid to evaluate the Rothe error, achieving tighter convergence of the Gaussian simulations would have been very time-consuming and was therefore not pursued. However, our results indicate that full convergence is, in principle, attainable.}

\section{Concluding remarks}\label{sec:Conclusion}

We have further developed Rothe’s method for the propagation of time-dependent, thawed Gaussians basis functions by applying it directly to orbital time-evolution equations, enabling accurate propagation of uncorrelated and correlated 1D multi-electron wave functions using TDHF and TDDFT in strong electric fields. Our results agree closely with reference grid calculations at intensities $I_0=10^{14} \WcmSq$ and $I_0=4\times 10^{14} \WcmSq$ for all methods and molecules considered when a sufficiently large number of Gaussians is used, with systematic improvements as the number of Gaussians is increased. In addition, we demonstrated that only a small number of Gaussians are sufficient to capture the most essential continuum effects for high-harmonic generation for the intensity $I_0=10^{14} \WcmSq$. Our findings indicate that Rothe’s method provides a flexible, computationally tractable route to strong-field many-particle time-dependent electronic structure calculations using Gaussian basis sets in 1D, offering a promising foundation for future extensions to more complex three-dimensional molecular systems.

Future work will focus on implementing the relevant three-dimensional Gaussian integrals analytically, alongside numerical techniques that avoid computing the kinetic part of the (squared) Hamiltonian. \cite{schrader2024time,schrader2025HeHe} Moreover, efficient parallelization of the integral evaluation enables a scalable implementation, significantly reducing computation time as the number of basis functions increases. This makes it possible to apply Rothe’s method to realistic three-dimensional molecular systems. In parallel, the formalism will be extended to correlated methods such as TDCC methods with time-dependent orbitals~\cite{kvaal2012ab, sato2018communication} and MCTDH, enabling treatment of correlated multi-electron dynamics.

A central challenge remains the repeated solution of similar, but difficult optimization problems at each time point. To make Rothe’s method competitive with grid-based approaches, developing a machine-learning algorithm that learns to optimize the Rothe error in as few iterations as possible, i.e., a \textit{learning to optimize} (L2O) method,~\cite{L2O_2016,L2O_primer_benchmark} is a possibility, which could speed up the underlying optimization. Such techniques have already shown to be promising in quantum circuit optimization. \cite{L2O_quantum} 

\section*{Data availability}
The data that supports the finding of this study is available on Zenodo, see Ref.~\citenum{Rothe_zenodo_TDHF}.

\section*{Author Declaration}
The authors have no conflicts to disclose.

\section*{Acknowledgments}
We thank Prof. Ludwik Adamowicz for helpful comments and discussions. The work was supported by the Research Council of Norway through its Centres of Excellence funding scheme, Project No. 262695. Some of the calculations were performed on resources provided by Sigma2, the National Infrastructure for High Performance Computing and Data Storage in Norway, Grant No. NN4654K.

\bibliography{main}

\end{document}